\documentclass[12pt]{spieman}  
\usepackage{amsmath,amsfonts,amssymb}
\usepackage{graphicx}
\usepackage{setspace}
\usepackage{tocloft}
\usepackage{xcolor}
\usepackage{placeins}
\usepackage{algorithm}
\usepackage{algpseudocode}

\newcommand{\argmin}{\operatornamewithlimits{argmin}}
\usepackage{booktabs}
\title{3D-DIP: Memory-Efficient Full-Volume Deep Image Prior for High-Resolution CT Reconstruction}

\author[a,*]{Ziyu Shu}
\author[a]{Xin Qian}
\author[b]{Zhixin Pan}
\author[a]{Junjie Wu}
\author[c]{Shuang Zhou}
\author[a]{Jingxuan Zhu}
\author[a]{Zhaozheng Yin}
\author[a]{Tiezhi Zhang}
\affil[a]{Stony Brook University, Radiation Oncology Department, Stony Brook, NY, US}
\affil[b]{Florida State University, Tallahassee, FL, US}
\affil[c]{Tetraimaging LLC, Stony Brook, NY, US}

\cftpagenumbersoff{figure}
\cftpagenumbersoff{table} 
\begin{document} 
\maketitle

\begin{abstract}

\textbf{Purpose:} Deep image prior (DIP) offers computed tomography (CT) reconstruction without external training data, but volumetric network optimization imposes substantial memory and computational demands. We aim to enable full-volume DIP reconstruction with low GPU-memory requirements and extend it to high-resolution sparse-view and limited-angle CT.

\textbf{Approach:} We first enable high-quality full-volume DIP reconstruction at $256\times256\times256$ using a compact 3D U-Net to reduce memory use and pipelined optimization to substantially accelerate DIP reconstruction by overlapping network computation with projection and backprojection. Building on this capability, the proposed 3D-DIP framework extends reconstruction to $512\times512\times512$ by upsampling the DIP estimate and fine-tuning it with adaptive steepest descent--projection onto convex sets (ASD-POCS). The two stages exploit complementary priors: DIP provides a nonlinear architectural prior that suppresses streak artifacts, while ASD-POCS combines the linear CT forward model with explicit regularization to refine local detail and reduce network-specific artifacts.

\textbf{Results:} At $256^3$, compact DIP reduces streak artifacts and improves signal-to-noise ratio (SNR) over ASD-POCS by 2.33--7.07~dB under sparse-view and limited-angle conditions in the tested brain and thoracic cases. Pipelined optimization achieves up to $2\times$ per-iteration acceleration without an observed loss of reconstruction quality. At $512^3$, the proposed 3D-DIP improves SNR over direct ASD-POCS by 3.13--3.83~dB under sparse-view acquisition and 6.91--7.57~dB under limited-angle acquisition. The complete framework uses approximately 18~GB of peak GPU memory. Reported timings imply a combined optimization time of approximately 13.3--13.7 minutes, adding 1.3--1.7 minutes to the approximately 12-minute ASD-POCS reconstruction.

\textbf{Conclusion:} 3D-DIP makes full-volume DIP reconstruction practical at $256^3$ and extends it to $512^3$ through complementary neural and iterative reconstruction priors. This approach lowers the computational barrier to high-resolution CT reconstruction without external training data, offering a memory-efficient route to improved image quality under sparse-view and limited-angle acquisition.
\end{abstract}


{\noindent \footnotesize\textbf{*}Ziyu Shu,  \linkable{zs919@nyu.edu} }

\begin{spacing}{2}   

\section{Introduction}
\label{sec:introduction}
Computed tomography (CT) reconstruction relies on analytical algorithms such as filtered backprojection (FBP) and iterative reconstruction (IR) methods that incorporate acquisition physics and image priors~\cite{fessler14,natterer,willemink2019evolution}. With incomplete or noisy measurements, these methods can produce streak artifacts, amplified noise, and loss of structural detail~\cite{sidky2006accurate,sidky2008image}. Sparse-view and limited-angle CT are particularly challenging because missing angular information leaves the reconstruction poorly constrained. Supervised deep learning can map projection data or preliminary reconstructions to higher-quality CT images and reduce artifacts and noise~\cite{jin2017deep,zhu2018image,wurfl2016deep}. Its performance, however, depends on representative training data and can vary across scanners, protocols, anatomies, and patient populations~\cite{bhadra2021hallucinations,antun2020instabilities,gottschling2023troublesome}. These challenges motivate reconstruction methods that provide an effective image prior without relying on external training data.

Deep image prior (DIP) uses an untrained convolutional network as an image representation and optimizes its parameters for each scan~\cite{ulyanov2018deep}. It requires no external training set, making it attractive when matched training data are scarce. Studies of two-dimensional CT have demonstrated its potential for sparse-view and limited-angle reconstruction~\cite{baguer2020computed,shu2022sparse,shu2024rbp}. Extending this approach to full CT volumes requires addressing the memory cost of three-dimensional feature maps and the repeated computation of network propagation, projection, and backprojection.

This study focuses on full-volume reconstruction, in which a 3D network represents the entire reconstruction grid. Such a representation provides spatial coupling across slices, but increasing the grid size rapidly raises the storage required for activations and gradients. Conventional network designs therefore limit the volume sizes that can be reconstructed within a modest GPU-memory budget.

We address this problem in two stages. First, we make direct volumetric DIP reconstruction practical at $256 \times 256 \times 256$ resolution. A compact 3D U-Net reduces the memory required for activations and gradients, while a pipelined optimization scheme overlaps network propagation with projection and backprojection to reduce iteration time. Second, we extend the reconstruction to a $512 \times 512 \times 512$ target grid. The $256^3$ DIP result is upsampled to initialize full-resolution IR, which uses the original projection data to refine measurement-supported details. This strategy also combines complementary priors: the nonlinear DIP prior suppresses artifacts caused by missing projections, while physics-based IR refinement improves data consistency and mitigates artifacts specific to the neural network.

The main contributions are as follows.
\begin{itemize}
\item We make direct full-volume DIP reconstruction practical at $256^3$ resolution through a compact 3D U-Net with reduced memory requirements.
\item We introduce a one-step delayed-gradient pipeline for DIP that overlaps network computation with projection and backprojection, reducing reconstruction time without an observed loss of image quality.
\item We extend DIP-guided reconstruction to $512^3$ by initializing full-resolution IR with the $256^3$ DIP result. This combines the complementary strengths of the nonlinear DIP prior and IR refinement without running a DIP network on the target grid.
\end{itemize}

The remainder of the paper is organized as follows. Section~\ref{sec:methods} presents the materials and methods, including the proposed 3D-DIP framework and experimental setup. Section~\ref{sec:exp} reports the results. Section~\ref{sec:discussion} discusses implications and limitations. Section~\ref{sec:conclusion} concludes the paper.

\subsection{Related Work}
\label{sec:related}
To address sparse-view and limited-angle CT, regularized IR combines the forward model with additional constraints, including sparsity, edge preservation, and nonlocal similarity~\cite{sidky2006accurate,chen2013limited,shu2025nonlocal,tian2011low,ritschl2011improved}. These hand-designed regularizers often promote sparse or smooth image representations, but may not sufficiently constrain highly ill-posed reconstructions when substantial angular information is missing. Supervised methods instead introduce data-driven priors through image-domain correction, projection completion, or learned mappings from measurements to images~\cite{wang2018image,jin2017deep,wurfl2016deep,lee2018deep,zhu2018image,xie2020deep}. By learning image structure from training data, they can achieve high reconstruction quality when test scans resemble the training distribution. However, their behavior can be unstable under changes in acquisition or anatomy, and learned patterns may appear in the reconstruction even when they are unsupported by the measurements~\cite{bhadra2021hallucinations,antun2020instabilities,gottschling2023troublesome}.

High-resolution 3D learned reconstruction also places substantial demands on GPU memory. LIRE extends learned reconstruction to cone-beam CT using invertible updates, patchwise computation, and CPU--GPU data transfer; its high-resolution model was fine-tuned on two NVIDIA A100 GPUs with 80~GB of memory each~\cite{moriakov2023lire}. DiffusionBlend forms a $256\times256\times500$ volume through a sequence of $256\times256$ two-dimensional slice reconstructions, using diffusion score blending to share information across neighboring slices~\cite{song2024diffusionblend}. Its slice-based parallel-beam measurement model differs from the fully three-dimensional coupling in cone-beam CT. The limited availability of representative volumetric training data with matched projections adds to these practical challenges~\cite{moriakov2023lire}.

These limitations motivate reconstruction with a prior optimized for each scan rather than pretrained on a population. DIP derives its regularization from network architecture and optimization rather than learned population statistics~\cite{ulyanov2018deep}. Local convolutions, parameter sharing, and multiscale representations favor spatially coherent images, and optimization often recovers dominant structure before fitting unstructured noise. This bias can help select plausible images among candidates that explain incomplete measurements. Baguer \textit{et al.} studied this behavior in tomography~\cite{baguer2020computed}, and later work applied DIP to sparse-view and limited-angle CT~\cite{shu2022sparse,shu2024rbp}.

Several extensions strengthen or modify this prior. Liu \textit{et al.} incorporated total variation (TV) into DIP~\cite{liu2019image}; Cui \textit{et al.} used auxiliary information in conditional DIP~\cite{cui2022unsupervised}; Shu and Pan introduced self-reinforcement~\cite{shu2024sdip}; and Nittscher \textit{et al.} restricted the learnable degrees of freedom~\cite{nittscher2024svddip}. Related developments include a masked DIP formulation for limited-angle SPECT~\cite{hori2025limited}. These studies address robustness and prior design, whereas the present work focuses on enabling full-volume optimization and high-resolution refinement within a limited memory budget.

Fully volumetric DIP has also been explored in emission tomography. Hashimoto \textit{et al.}, for example, used block-iterative optimization for DIP-based PET reconstruction on $128\times128\times64$ volumes~\cite{hashimoto2023fully3d}. However, high-resolution 3D DIP reconstruction remains limited by GPU memory requirements. Our focus is efficient DIP reconstruction on a $256^3$ CT grid and its use as a structural prior for a $512^3$ reconstruction.

\section{Materials and Methods}
\label{sec:methods}
Let $\boldsymbol{y}$ denote the measured projection data, $\boldsymbol{H}$ the CT forward-projection operator, and $\boldsymbol{H}^{T}$ the corresponding backprojection operator. In DIP reconstruction, the image volume is represented by a DIP network $G_{\boldsymbol{w}}(\boldsymbol{z})$, where $\boldsymbol{z}$ is a fixed input tensor, typically random noise, and $\boldsymbol{w}$ denotes the trainable network parameters. For a given scan, DIP reconstructs the image by optimizing $\boldsymbol{w}$ directly on the measured data, rather than by pretraining on an external dataset. The basic DIP reconstruction problem can be written as
\begin{equation}
\boldsymbol{w}^{\star} = \argmin_{\boldsymbol{w}} \left\| \boldsymbol{y} - \boldsymbol{H} G_{\boldsymbol{w}}(\boldsymbol{z}) \right\|_2^2.
\label{eq:dip}
\end{equation}
For high-resolution volumetric CT, solving~\eqref{eq:dip} directly with a standard 3D U-Net can exceed the available GPU memory. We first simplify the network and overlap network computation with projector operations, making efficient full-volume DIP reconstruction at $256 \times 256 \times 256$ practical. Building on this reconstruction, we combine the nonlinear DIP prior with explicit physics-based IR regularization to achieve accurate and computationally efficient $512 \times 512 \times 512$ reconstruction with low GPU-memory requirements. The following subsections describe the compact network, pipelined optimization, low-to-high-resolution strategy, complete reconstruction workflow, and experimental setup.

\subsection{Memory-Efficient Network Design}
\label{sec:netsimpli}
The memory cost of 3D DIP is dominated by volumetric tensors used for network evaluation and differentiation. In particular, backpropagation requires intermediate activations from the forward pass to be retained in GPU memory. A U-Net with many resolution levels and channels therefore stores large three-dimensional activation maps throughout its encoder and decoder~\cite{ronneberger2015u}. Reducing network depth and width decreases both the number and size of these stored activations, thereby lowering the memory required for full-volume optimization.

Our approach matches network capacity to the role of DIP in reconstruction. A DIP network is optimized for a single scan and serves as a structured parameterization of the image, so architectures developed for supervised learning need not provide the best balance between reconstruction quality and computational cost. Previous studies have shown that constraining untrained networks can limit their ability to fit noise or spurious components~\cite{heckel2019deepdecoder,jo2021rethinking,wang2023early,nittscher2024svddip}. These observations motivate an empirical evaluation of smaller 3D networks.

We vary the U-Net depth and initial channel count to examine the trade-off between reconstruction quality and computational cost. Here, network depth denotes the number of encoder--decoder levels, and the initial channel count denotes the number of feature channels at the first level. Each encoder or decoder block contains two $3\times3\times3$ convolutions, each followed by batch normalization and a leaky rectified linear unit. Downsampling uses a stride-2 convolution, upsampling uses trilinear interpolation followed by a $3\times3\times3$ convolution, and corresponding encoder and decoder features are joined by skip connections. Section~\ref{sec:exp-ustru} reports a systematic evaluation of network depth and channel count. Based on this evaluation, we select a four-level U-Net with channel counts of 32, 64, 128, and 256 for the $256^3$ DIP stage and for the controlled comparison across reconstruction grids.

\subsection{Pipelined DIP Optimization}
\label{sec:pipeline}
\begin{figure}[htbp]
    \begin{minipage}[b]{\linewidth}
        \centering
        \includegraphics[width=\linewidth]{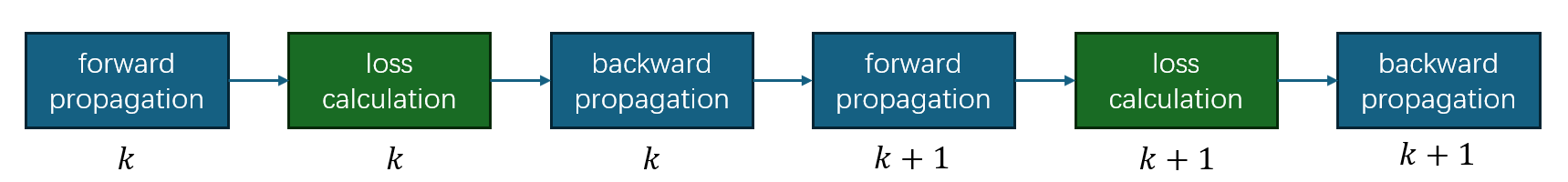}
        \centerline{(a)}\medskip
    \end{minipage}

    \begin{minipage}[b]{\linewidth}
        \centering
        \includegraphics[width=\linewidth]{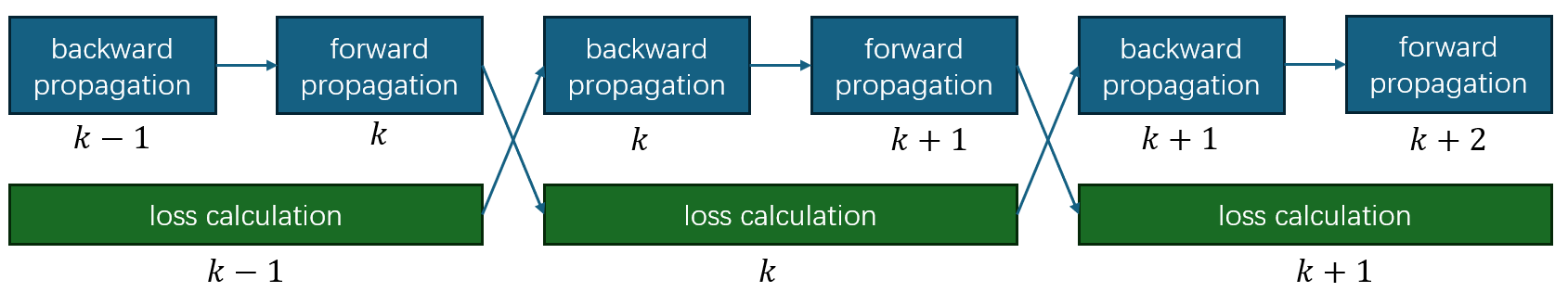}
        \centerline{(b)}\medskip
    \end{minipage}
    \caption{Comparison between conventional DIP optimization and the proposed pipelined optimization. (a) In the conventional workflow, network forward/backward propagation and CT loss evaluation are executed sequentially in each iteration. (b) In the proposed pipeline, network propagation and projector operations are overlapped across consecutive iterations, reducing the effective iteration time.}
    \label{fig:pipeline}
\end{figure}

Conventional DIP optimization executes network computation and CT projection and backprojection sequentially. Because a three-dimensional system matrix is too large to store explicitly, projection and backprojection must be evaluated on the fly in every iteration and can account for a substantial fraction of the reconstruction time. The proposed pipeline overlaps these projector operations with network computation, hiding much of their cost and permitting an ideal speedup of up to $2\times$ when the two workloads are comparable.

Existing projector implementations, including ray-integration lookup tables, box-spline models, separable footprints, and blob-based methods, generally balance computational speed against numerical accuracy~\cite{ha2017look,shu2020gram,shu2022exact,zhang2019box,long20103d,ziegler2006blob}. Pipelining relaxes this trade-off at the workflow level: a more accurate but relatively expensive projector can be used with little increase in effective iteration time, provided that its cost does not exceed the network computation available for overlap and that synchronization overhead remains small.

The pipeline introduces a one-iteration delay in the image-domain gradient. At iteration $k$, a forward pass produces $\boldsymbol{x}_k=G_{\boldsymbol{w}_k}(\boldsymbol{z})$. The projector computes $\boldsymbol{g}_k=2\boldsymbol{H}^{T}(\boldsymbol{H}\boldsymbol{x}_k-\boldsymbol{y})$ for the next update, while network backpropagation uses the buffered gradient $\boldsymbol{g}_{k-1}$. The parameter update therefore uses the current network Jacobian with the previous image-domain gradient. The first iteration is executed serially to initialize the gradient buffer. This schedule retains the data-fidelity objective and projector evaluations but changes when the image-domain gradient is used, allowing projector work to overlap with network backpropagation and optimization. Figure~\ref{fig:pipeline} illustrates the serial and pipelined schedules.

Our implementation uses two separate non-default CUDA streams for network and projector operations, with CUDA events enforcing dependencies. The projector stream waits for the current network output, and the network stream waits for the delayed gradient needed by its next update. Separate buffers prevent either stream from overwriting an image or gradient still in use. This scheduling permits independent operations to overlap without a device-wide barrier at every exchange.

If the network propagation cost is denoted by $T_{\text{net}}$ and the projector/backprojector cost by $T_{\text{proj}}$, then the conventional serial implementation requires approximately
\begin{equation}
T_{\text{serial}} \approx T_{\text{net}} + T_{\text{proj}}
\end{equation}
per iteration. In contrast, under our proposed overlapped execution, the pipelined scheme reduces the effective iteration cost to approximately
\begin{equation}
T_{\text{pipe}} \approx \max \left( T_{\text{net}}, T_{\text{proj}} \right)+T_{\text{sync}},
\end{equation}
where $T_{\text{sync}}$ represents the effective overhead beyond ideal overlap, including synchronization, scheduling, resource contention, and dependency-induced idle time. This is an idealized steady-state model: the current network output must be available before its projection can begin, so the attainable overlap also depends on the division between network forward and backward propagation. A speedup approaching $2\times$ requires similar component costs and nearly complete overlap with little overhead.

We evaluate both the realized speedup and the effect of the delayed update on reconstruction quality. The latter is important because DIP can fit noise or model mismatch after recovering the dominant structure~\cite{ulyanov2018deep,cascarano2022first}. A one-step delay perturbs the optimization trajectory and may affect this behavior. Section~\ref{sec:pipeline-performance} reports the measured runtime, loss, and SNR trajectories, and Section~\ref{sec:discussion} considers a possible interpretation.

\subsection{Low-Resolution DIP-Guided Reconstruction}
\label{sec:low-resolution-guidance}

The compact network and pipeline enable DIP optimization on a $256^3$ grid. Directly extending the same network to $512^3$ would increase the number of voxels, and therefore the storage required by every volume-sized feature map, by a factor of eight. To avoid this increase, we transfer the $256^3$ DIP estimate to the target grid by trilinear interpolation and refine it with ASD-POCS. This low-to-high-resolution strategy retains the structural information recovered by DIP without allocating network activations on the $512^3$ grid.

This strategy rests on two considerations. First, we hypothesize that the artifact-suppression benefit of DIP persists at reduced resolution, allowing a coarse DIP reconstruction to provide a suitable structural estimate for initializing the fine grid. Section~\ref{sec:resolution-extension} evaluates this hypothesis using a fixed network architecture across different grid sizes.

Second, DIP and ASD-POCS provide complementary priors for the ill-posed reconstruction. DIP imposes a highly nonlinear architectural prior that favors spatially coherent structure and suppresses streak artifacts caused by missing projections~\cite{ulyanov2018deep,heckel2019deepdecoder,jo2021rethinking,wang2023early,nittscher2024svddip}, although network optimization can introduce neural-network-specific artifacts. ASD-POCS, by comparison, applies explicit image-domain regularization based on the CT forward model, nonnegativity, and TV reduction. Direct ASD-POCS cannot fully eliminate missing-projection artifacts, but its stable physics-based updates can attenuate artifacts introduced by the DIP network.

Together, these considerations motivate using the upsampled DIP reconstruction as an artifact-suppressed structural initialization and ASD-POCS to refine measurement-supported details and suppress network-specific artifacts. The DIP prior is transferred through initialization, without an additional penalty that forces the refined image to remain close to the network output. The resulting combination of a nonlinear architectural prior and explicit physics-based regularization differs from conventional multiresolution IR, which applies essentially the same reconstruction model successively across resolutions~\cite{cao2016multiresolution}. Section~\ref{sec:discussion} further discusses the complementary roles of the two priors and their distinction from conventional multiresolution IR.

\subsection{Overall Reconstruction Workflow}
\label{sec:workflow}
\begin{figure*}[t]
    \centering
    \includegraphics[width=\linewidth]{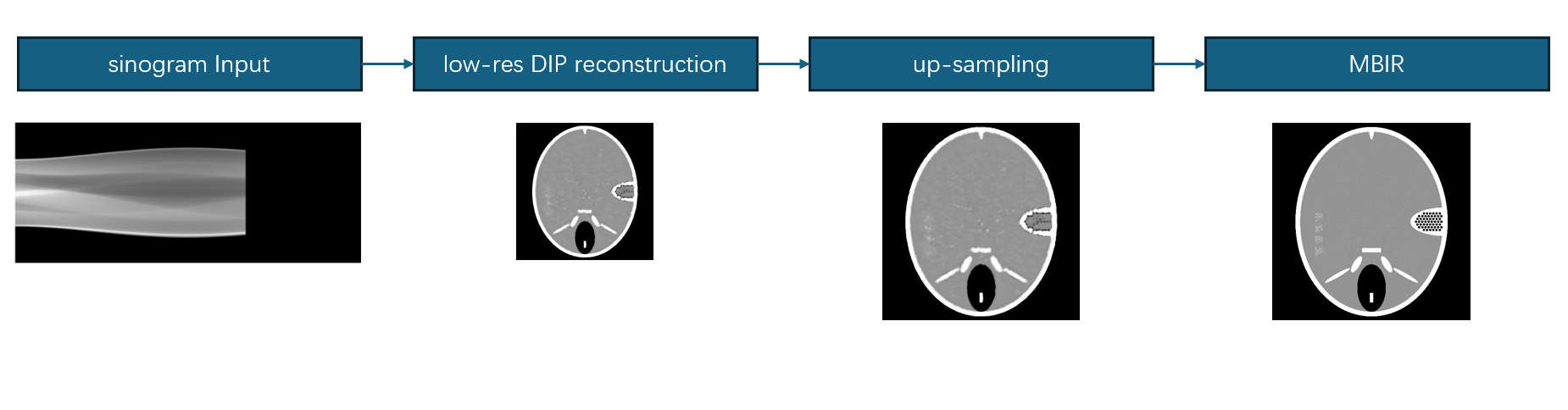}
    \caption{Overview of the proposed 3D-DIP framework. Pipelined DIP reconstruction is first performed at $256 \times 256 \times 256$ resolution to obtain an artifact-suppressed structural estimate. The result is then upsampled to the $512 \times 512 \times 512$ target grid and used to initialize full-resolution ASD-POCS refinement.}
    \label{fig:workflow}
\end{figure*}

Figure~\ref{fig:workflow} summarizes the two stages. The reduced DIP grid has $256 \times 256 \times 256$ voxels at $1.0 \times 1.0 \times 1.0~\mathrm{mm}^3$, and the target grid has $512 \times 512 \times 512$ voxels at $0.5 \times 0.5 \times 0.5~\mathrm{mm}^3$. Both cover the same physical field of view. The detector sampling, projection angles, and scanner geometry are unchanged. The reduced-resolution DIP stage uses the selected four-level 3D U-Net with 32 initial channels. On the reduced grid, the DIP objective is
\begin{equation}
\min_{\boldsymbol{w}}
\left\|
\boldsymbol{y} -
\boldsymbol{H}_{\mathrm{lr}}G_{\boldsymbol{w}}(\boldsymbol{z})
\right\|_2^2,
\end{equation}
where $\boldsymbol{y}$ is the original projection vector and $\boldsymbol{H}_{\mathrm{lr}}$ is the forward operator for the reduced grid. After optimization, the resulting parameters $\widehat{\boldsymbol{w}}_{\mathrm{lr}}$ define the full-resolution initial image
\begin{equation}
\boldsymbol{x}^{(0)} =
\mathcal{U}\left(G_{\widehat{\boldsymbol{w}}_{\mathrm{lr}}}(\boldsymbol{z})\right),
\end{equation}
where $\mathcal{U}:\mathbb{R}^{256\times256\times256}\rightarrow\mathbb{R}^{512\times512\times512}$ denotes trilinear interpolation over the same physical field of view. The resulting $\boldsymbol{x}^{(0)}$ transfers the DIP structural estimate to the target grid but does not itself generate high-resolution detail. That detail is recovered in the second stage using the original projection data and the high-resolution system model.

All three-dimensional IR experiments in this work use adaptive steepest descent--projection onto convex sets (ASD-POCS)~\cite{sidky2008image}. This algorithm directly updates the image volume, rather than the DIP network parameters, by alternating data-consistency projection steps with adaptive steepest-descent TV reduction. At iteration $k$, the high-resolution data residual is
\begin{equation}
\boldsymbol{r}^{(k)}=\boldsymbol{y}-\boldsymbol{H}_{\mathrm{hr}}\boldsymbol{x}^{(k)},
\end{equation}
where $\boldsymbol{H}_{\mathrm{hr}}$ is the system operator defined on the target grid. The POCS updates use this residual to recover details supported by the projection data, while the TV steps suppress unstable variations caused by incomplete sampling.

We use the parameter settings reported by Sidky and Pan: the initial algebraic-reconstruction relaxation parameter is $\beta_0=1.0$, its reduction factor is $\beta_{\mathrm{red}}=0.995$, the number of TV gradient-descent steps per outer iteration is $n_g=20$, the initial TV step-size ratio is $\alpha=0.2$, the maximum ratio between the TV-descent and POCS update magnitudes is $r_{\max}=0.95$, and the TV step-size reduction factor is $\alpha_{\mathrm{red}}=0.95$. Nonnegativity is enforced after each POCS update.

In the $512^3$ comparisons reported in Sections~\ref{sec:resolution-extension} and~\ref{sec:clinical-volumes}, the direct ASD-POCS baseline starts from a zero volume, whereas 3D-DIP starts from $\boldsymbol{x}^{(0)}$; both use identical ASD-POCS update parameters and iteration budgets, as specified in Section~\ref{sec:experimental-setup}. Consequently, these comparisons isolate the effect of the DIP-derived initialization: the first stage supplies an artifact-suppressed structural candidate, and the second-stage physics-based optimization refines it at the target resolution.

\subsection{Experimental Setup}
\label{sec:experimental-setup}
The full-resolution experiments use cone-beam CT geometry with a $512\times512\times512$ reconstruction grid and $0.5\times0.5\times0.5~\mathrm{mm}^3$ voxels. The detector pixel size is $0.388\times0.388~\mathrm{mm}^2$, the source-to-detector distance is 150~cm, and the source-to-axis distance is 50~cm. A custom ray-tracing implementation provides projection and backprojection with detector blur modeling~\cite{shu2020gram,shu2022exact}. For the simulated measurements, the incident photon count is set to $N_0=10^5$ to represent moderate quantum noise. Given a line integral $p$, the detected photon count is sampled from $N\sim\operatorname{Poisson}(N_0e^{-p})$, and the noisy projection is obtained by the negative logarithmic transform. Reference volumes are upsampled by a factor of two before projection simulation to reduce the use of identical discretizations for data generation and reconstruction.

Reconstruction quality is evaluated using signal-to-noise ratio (SNR) and the structural similarity index measure (SSIM). SNR measures global reconstruction error relative to the reference signal energy, while SSIM provides a complementary measure of structural fidelity. We report SNR across acquisition conditions and include SSIM in representative reconstruction comparisons.

Experiments were performed on an NVIDIA GeForce RTX~4090 with 24~GB of device memory. Networks are implemented in PyTorch. Unless otherwise stated, RMSProp uses an initial learning rate of $10^{-4}$, multiplied by 0.9 every 250 iterations. All DIP reconstructions use 2500 iterations. ASD-POCS uses 2500 outer iterations in the direct $256^3$ and resolution comparisons, and 1000 outer iterations for both direct and DIP-initialized $512^3$ reconstruction. Metrics are evaluated at the final iteration; reference-based metrics are not used to select the stopping point.

\section{Results}
\label{sec:exp}
We first evaluate the compact network and delayed-gradient pipeline, then assess reconstruction at $256^3$ and its extension to $512^3$. Experiments over a range of angular sampling conditions examine the complete framework, followed by a summary of memory use and runtime.

\subsection{Architecture Simplification}
\label{sec:exp-ustru}
\begin{figure}[htbp]
   \begin{minipage}[b]{0.3\linewidth}
        \centering
        \includegraphics[width=\linewidth]{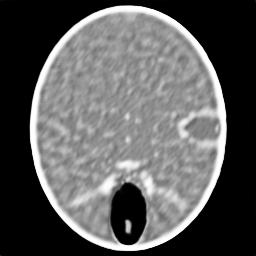}
        \centerline{8}
    \end{minipage}
    \begin{minipage}[b]{0.3\linewidth}
        \centering
        \includegraphics[width=\linewidth]{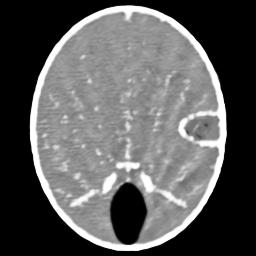}
        \centerline{16}
    \end{minipage}
    \begin{minipage}[b]{0.3\linewidth}
        \centering
        \includegraphics[width=\linewidth]{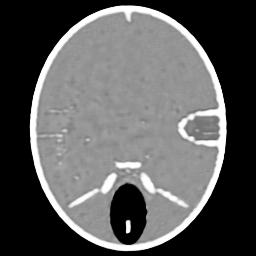}
        \centerline{32}
    \end{minipage}

    \vspace{0.2cm}
    \begin{minipage}[b]{0.3\linewidth}
        \centering
        \includegraphics[width=\linewidth]{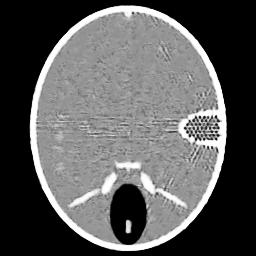}
        \centerline{64}
    \end{minipage}
    \begin{minipage}[b]{0.3\linewidth}
        \centering
        \includegraphics[width=\linewidth]{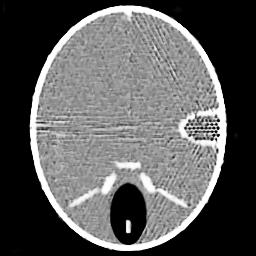}
        \centerline{128}
    \end{minipage}
    \begin{minipage}[b]{0.3\linewidth}
        \centering
        \includegraphics[width=\linewidth]{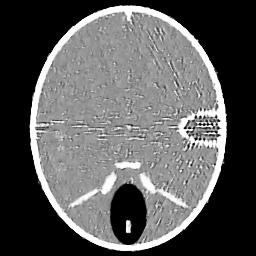}
        \centerline{256}
    \end{minipage}
    \caption{Representative DIP reconstructions obtained by varying the channel width of a five-level 3D U-Net under incomplete sampling. The standard 64-channel network uses 64, 128, 256, 512, and 1024 channels from the finest to the coarsest level. All other configurations preserve the same network depth, convolutional blocks, and skip connections while scaling the channels at every level in proportion to the indicated initial channel count; for example, the 32-channel configuration uses 32, 64, 128, 256, and 512 channels. The number below each image denotes the initial channel count. The nonmonotonic visual changes indicate that increasing network width does not necessarily improve DIP reconstruction.}
    \label{fig:unetstru}
\end{figure}

We first examine the trade-off between network capacity and reconstruction quality. We take a standard five-level U-Net with 64 initial channels as the reference configuration. Although this network provides substantial representation capacity, DIP optimizes a single volume and may not require the capacity commonly used in supervised learning. Indeed, increasing network capacity does not necessarily improve DIP reconstruction, as illustrated by the nonmonotonic changes with network width in Fig.~\ref{fig:unetstru}. Compact networks have also been effective in other tomographic DIP applications~\cite{gong2018pet}.

We evaluate U-Nets with different depths and initial channel counts on four representative 3D objects: a FORBILD phantom~\cite{yu2012simulation}, a random ellipsoid phantom based on ellipsoidal CT phantom models~\cite{shepp1974fourier}, a brain CT volume~\cite{chilamkurthy2018deep}, and a lung CT volume~\cite{armato2011lung}. Reconstruction quality is summarized relative to a five-level U-Net with 64 initial channels, which serves as the baseline.

\begin{table}[htbp]
\centering
\begin{tabular}{l|cccc}
\toprule
\textbf{Initial} & \multicolumn{4}{c}{\textbf{Resolution levels}} \\
\cmidrule(lr){2-5}
\textbf{channels} & 3 & 4 & 5 & 6 \\
\midrule
32 & -0.92 & \textbf{+0.45} & +0.15 & -0.98 \\
48 & -0.59 & +0.43 & +0.09 & -1.21 \\
64 & -0.21 & +0.43 & 0.00 & -1.42 \\
80 & -0.23 & +0.42 & -0.21 & -1.62 \\
96 & +0.11 & +0.23 & -0.53 & -1.93 \\
\bottomrule
\end{tabular}
\vspace{4pt}
\caption{Mean SNR difference (dB) across four test objects relative to the five-level network with 64 initial channels. Positive values favor the tested configuration.}
\label{tab:level_channel}
\end{table}

The four-level, 32-channel network gives the largest mean SNR increase over the reference configuration, 0.45~dB (Table~\ref{tab:level_channel}). We therefore select this configuration for the full-volume $256^3$ stage. Its reconstruction quality is evaluated directly in Section~\ref{sec:volumetric256}.

\subsection{Pipelined Optimization Performance}
\label{sec:pipeline-performance}
We compare serial and pipelined DIP optimization to evaluate two aspects of the delayed update: the reduction in per-iteration runtime and its effect on loss and reconstruction SNR.

Table~\ref{tab:pipeline_time} reports average per-iteration wall-clock times at two volume sizes. Component times were measured separately, and serial and pipeline times were measured over their complete optimization loops. Increasing the projection count raises the projector cost while leaving the network cost approximately unchanged. The reported speedups range from $1.28\times$ to $2.00\times$, with the greatest benefit when network and projector costs are similar.

\begin{table}[htbp]
\centering
\small
\setlength{\tabcolsep}{3pt}
\begin{tabular}{llccccc}
\toprule
\textbf{Proj. no.} & \textbf{Volume size} & \textbf{Proj./backproj.} & \textbf{Network} & \textbf{Serial} & \textbf{Pipeline} & \textbf{Speedup} \\
 & & \textbf{time (s)} & \textbf{time (s)} & \textbf{time (s)} & \textbf{time (s)} & \textbf{(Serial/Pipeline)}\\
\midrule
30 & $256 \times 256 \times 256$ & 0.01 & 0.03 & 0.04 & 0.03 & $1.33\times$\\
30 & $512 \times 512 \times 512$ & 0.06 & 0.22 & 0.28 & 0.22 & $1.28\times$ \\
60 & $256 \times 256 \times 256$ & 0.02 & 0.03 & 0.05 & 0.03 & $1.67\times$\\
60 & $512 \times 512 \times 512$ & 0.12 & 0.22 & 0.34 & 0.22 & $1.55\times$ \\
90 & $256 \times 256 \times 256$ & 0.03 & 0.03 & 0.06 & 0.03 & $2.00\times$\\
90 & $512 \times 512 \times 512$ & 0.19 & 0.22 & 0.41 & 0.22 & $1.86\times$ \\
120 & $256 \times 256 \times 256$ & 0.04 & 0.03 & 0.07 & 0.04 & $1.75\times$\\
120 & $512 \times 512 \times 512$ & 0.25 & 0.22 & 0.47 & 0.25 & $1.88\times$ \\
\bottomrule
\end{tabular}
\vspace{4pt}
\caption{Mean per-iteration wall-clock times and speedups. Component times were measured separately; serial and pipeline times cover their respective optimization loops.}
\label{tab:pipeline_time}
\end{table}

\begin{figure}[htbp]
   \begin{minipage}[b]{0.88\linewidth}
        \centering
        \includegraphics[width=\linewidth]{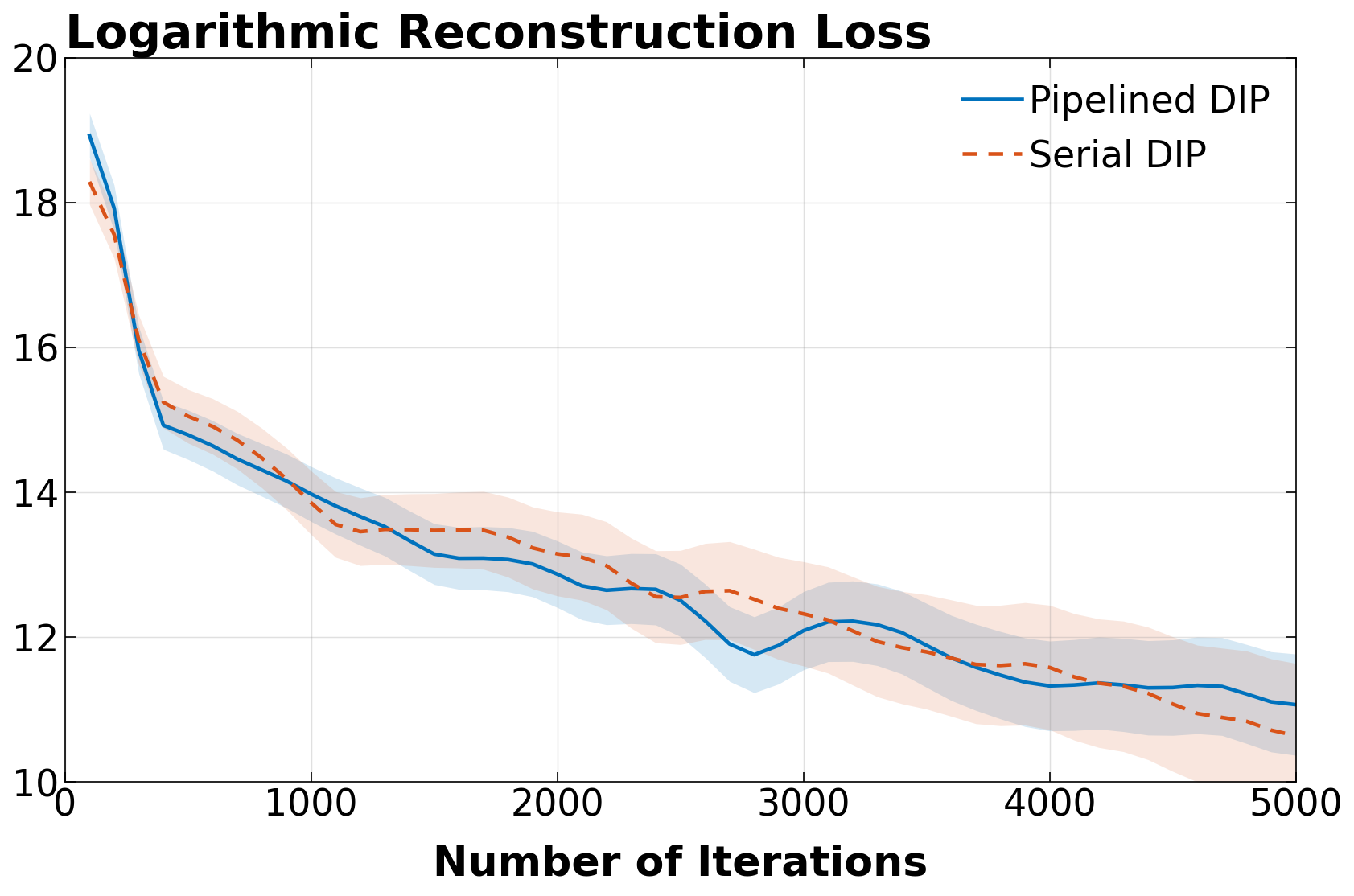}
        \centerline{(a) Average reconstruction loss}
    \end{minipage}

    \vspace{0.2cm}
    \begin{minipage}[b]{0.88\linewidth}
        \centering
        \includegraphics[width=\linewidth]{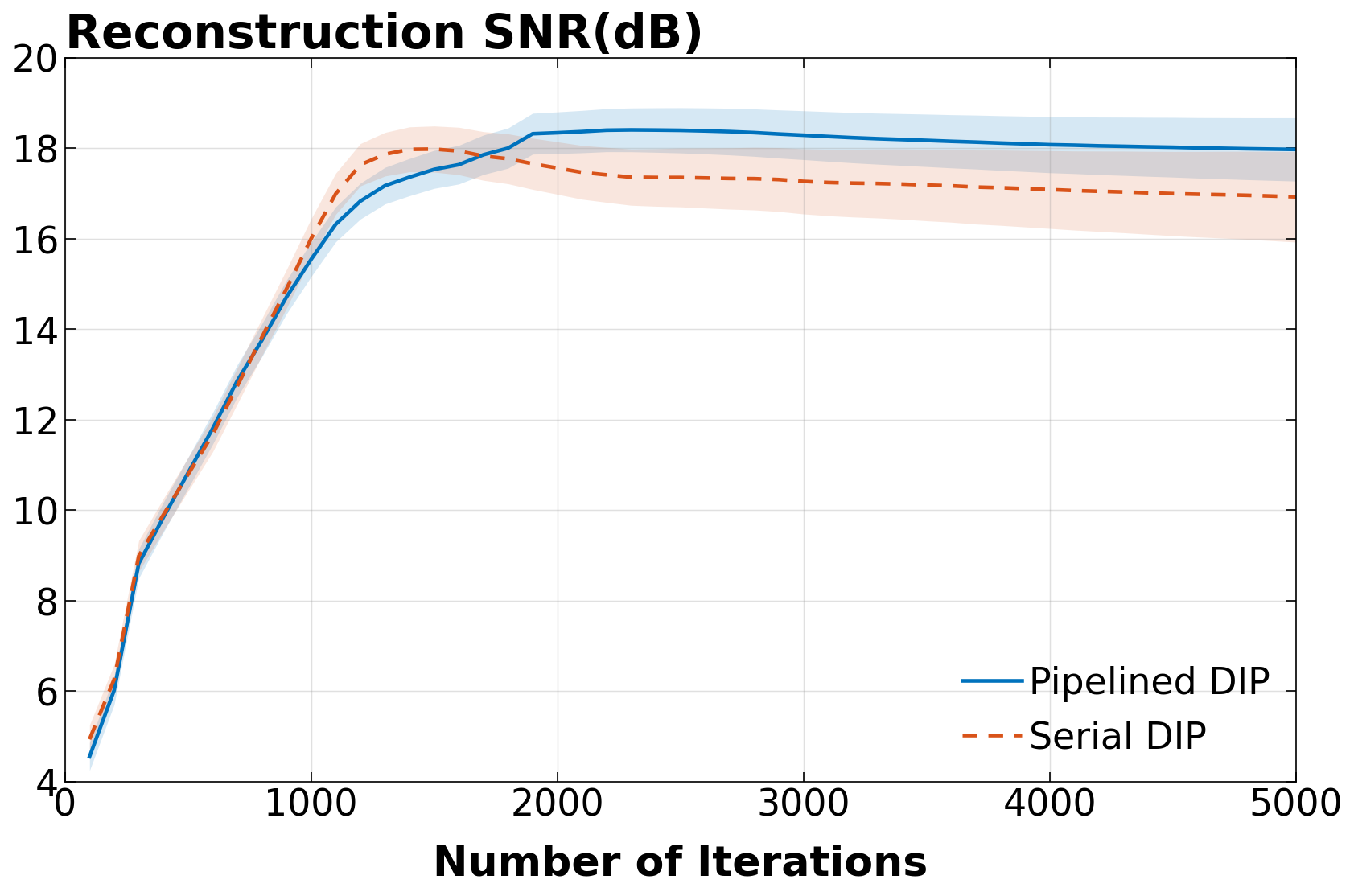}
        \centerline{(b) Average reconstruction SNR}
    \end{minipage}
    \caption{Error-band plots of serial and pipelined DIP optimization for the FORBILD phantom with 30 sparse-view projections over $360^\circ$. The central curves show the mean logarithmic reconstruction loss (a) and mean reconstruction SNR (b), and the shaded regions show the observed ranges across repeated reconstructions. The methods exhibit similar loss reduction, while pipelined DIP attains a slightly higher final SNR.}
    \label{fig:pipeline_curves}
\end{figure}

We further compare serial and pipelined optimization on the same four objects used in the architecture study under both sparse-view and limited-angle acquisition. The sparse-view setting uses 30 projections over $360^\circ$, and the limited-angle setting uses 120 projections over a $120^\circ$ angular range. Across these experiments, the one-step gradient delay does not materially degrade convergence behavior or reconstruction accuracy. The loss trajectories remain similar, and the final reconstruction SNR of pipelined DIP is slightly higher than that of serial DIP in our experiments. Figure~\ref{fig:pipeline_curves} presents representative error-band plots for the FORBILD phantom under the 30-projection sparse-view setting. The central loss curves are nearly indistinguishable, with substantial overlap between their error bands. Serial DIP reaches its peak SNR earlier, whereas the subsequent decrease is less pronounced for pipelined DIP. The other tested objects and acquisition conditions exhibit similar optimization behavior. Together, the timing and image-quality results show that the one-step delayed update accelerates DIP reconstruction without reducing reconstruction accuracy and can provide a small improvement in the final reconstruction.

\subsection{Volumetric DIP Reconstruction at \texorpdfstring{$256^3$}{256 cubed}}
\label{sec:volumetric256}

\begin{figure*}[htbp]
\centering
\begin{minipage}[b]{0.19\linewidth}
\centering
\includegraphics[width=\linewidth]{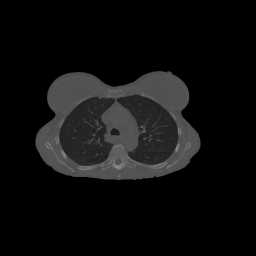}
\end{minipage}%
\hfill
\begin{minipage}[b]{0.19\linewidth}
\centering
\includegraphics[width=\linewidth]{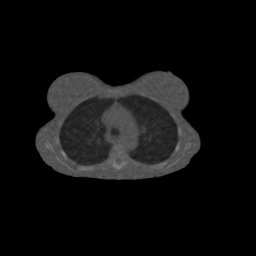}
\end{minipage}%
\hfill
\begin{minipage}[b]{0.19\linewidth}
\centering
\includegraphics[width=\linewidth]{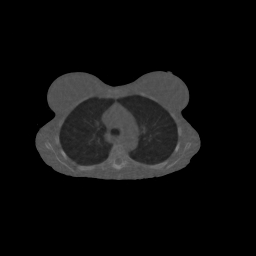}
\end{minipage}%
\hfill
\begin{minipage}[b]{0.19\linewidth}
\centering
\includegraphics[width=\linewidth]{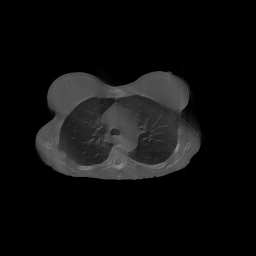}
\end{minipage}%
\hfill
\begin{minipage}[b]{0.19\linewidth}
\centering
\includegraphics[width=\linewidth]{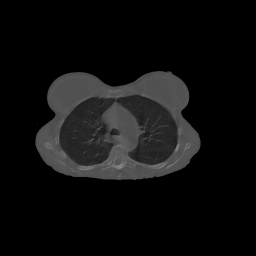}
\end{minipage}%
\vspace{0.12cm}
\begin{minipage}[b]{0.19\linewidth}
\centering
\includegraphics[width=\linewidth]{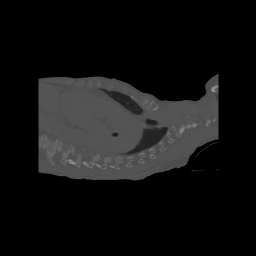}
\end{minipage}%
\hfill
\begin{minipage}[b]{0.19\linewidth}
\centering
\includegraphics[width=\linewidth]{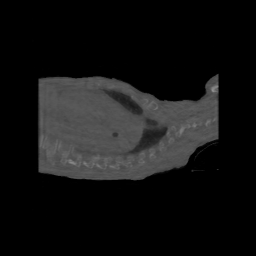}
\end{minipage}%
\hfill
\begin{minipage}[b]{0.19\linewidth}
\centering
\includegraphics[width=\linewidth]{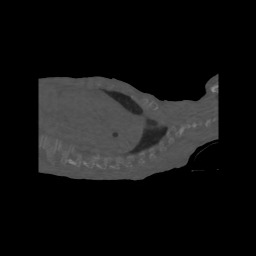}
\end{minipage}%
\hfill
\begin{minipage}[b]{0.19\linewidth}
\centering
\includegraphics[width=\linewidth]{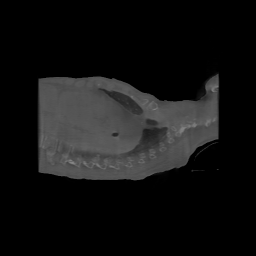}
\end{minipage}%
\hfill
\begin{minipage}[b]{0.19\linewidth}
\centering
\includegraphics[width=\linewidth]{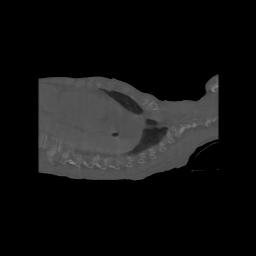}
\end{minipage}%
\vspace{0.12cm}
\begin{minipage}[b]{0.19\linewidth}
\centering
\includegraphics[width=\linewidth]{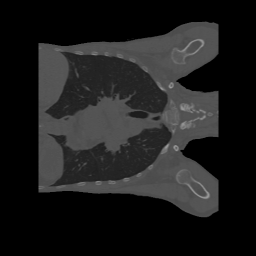}
\centerline{\scriptsize Reference}
\centerline{\scriptsize SNR/SSIM}
\end{minipage}%
\hfill
\begin{minipage}[b]{0.19\linewidth}
\centering
\includegraphics[width=\linewidth]{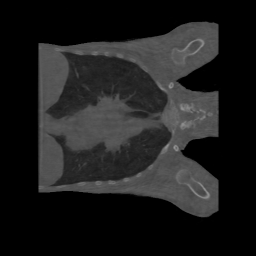}
\centerline{\scriptsize Few-view IR}
\centerline{\scriptsize 24.59~dB/0.9853}
\end{minipage}%
\hfill
\begin{minipage}[b]{0.19\linewidth}
\centering
\includegraphics[width=\linewidth]{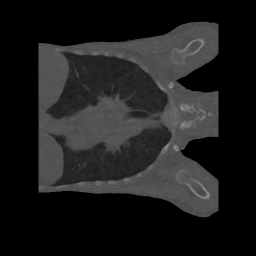}
\centerline{\scriptsize Few-view DIP}
\centerline{\scriptsize 26.92~dB/0.9930}
\end{minipage}%
\hfill
\begin{minipage}[b]{0.19\linewidth}
\centering
\includegraphics[width=\linewidth]{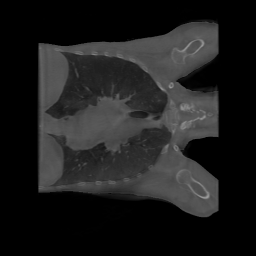}
\centerline{\scriptsize Limited-angle IR}
\centerline{\scriptsize 19.39~dB/0.9781}
\end{minipage}%
\hfill
\begin{minipage}[b]{0.19\linewidth}
\centering
\includegraphics[width=\linewidth]{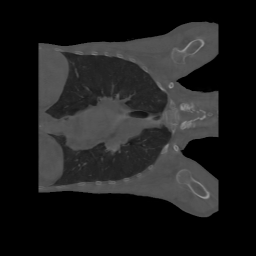}
\centerline{\scriptsize Limited-angle DIP}
\centerline{\scriptsize 26.46~dB/0.9951}
\end{minipage}%
\caption{Representative $256^3$ reconstruction results for the chest CT volume. The sparse-view setting uses 30 projections uniformly distributed over $360^\circ$; the limited-angle setting uses 120 projections acquired at $1^\circ$ intervals over $120^\circ$. Rows show axial, coronal, and sagittal slices; columns show the reference, sparse-view ASD-POCS, sparse-view DIP, limited-angle ASD-POCS, and limited-angle DIP. Each column is labeled below the third row, with reconstruction values reported as SNR in dB/SSIM.}
\label{fig:clinical256_chest}
\end{figure*}

\begin{figure*}[htbp]
\centering
\begin{minipage}[b]{0.19\linewidth}
\centering
\includegraphics[width=\linewidth]{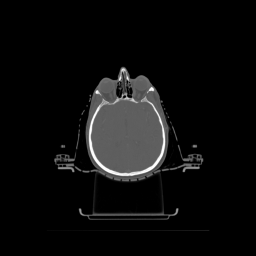}
\end{minipage}%
\hfill
\begin{minipage}[b]{0.19\linewidth}
\centering
\includegraphics[width=\linewidth]{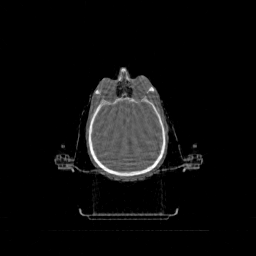}
\end{minipage}%
\hfill
\begin{minipage}[b]{0.19\linewidth}
\centering
\includegraphics[width=\linewidth]{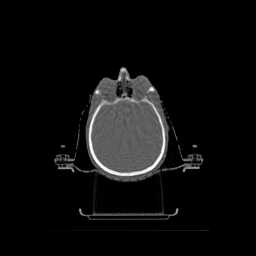}
\end{minipage}%
\hfill
\begin{minipage}[b]{0.19\linewidth}
\centering
\includegraphics[width=\linewidth]{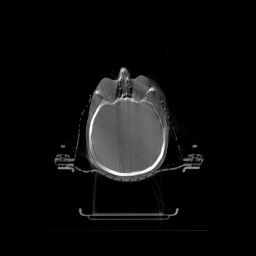}
\end{minipage}%
\hfill
\begin{minipage}[b]{0.19\linewidth}
\centering
\includegraphics[width=\linewidth]{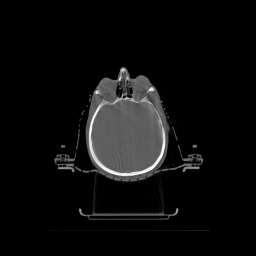}
\end{minipage}%
\vspace{0.12cm}
\begin{minipage}[b]{0.19\linewidth}
\centering
\includegraphics[width=\linewidth]{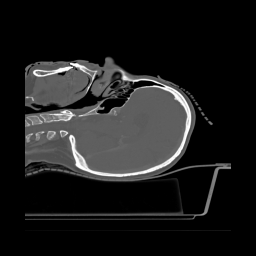}
\end{minipage}%
\hfill
\begin{minipage}[b]{0.19\linewidth}
\centering
\includegraphics[width=\linewidth]{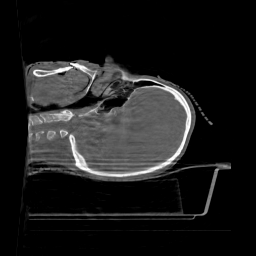}
\end{minipage}%
\hfill
\begin{minipage}[b]{0.19\linewidth}
\centering
\includegraphics[width=\linewidth]{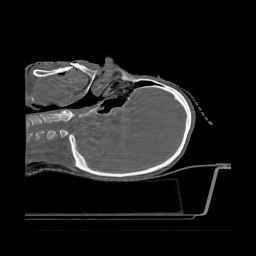}
\end{minipage}%
\hfill
\begin{minipage}[b]{0.19\linewidth}
\centering
\includegraphics[width=\linewidth]{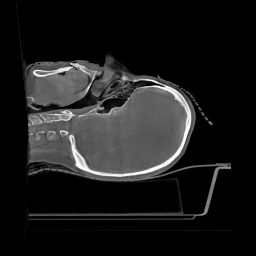}
\end{minipage}%
\hfill
\begin{minipage}[b]{0.19\linewidth}
\centering
\includegraphics[width=\linewidth]{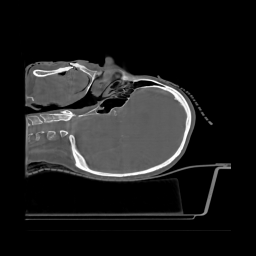}
\end{minipage}%
\vspace{0.12cm}
\begin{minipage}[b]{0.19\linewidth}
\centering
\includegraphics[width=\linewidth]{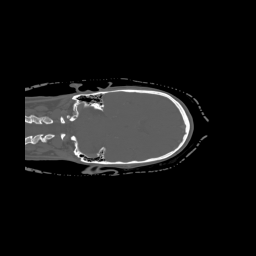}
\centerline{\scriptsize Reference}
\centerline{\scriptsize SNR/SSIM}
\end{minipage}%
\hfill
\begin{minipage}[b]{0.19\linewidth}
\centering
\includegraphics[width=\linewidth]{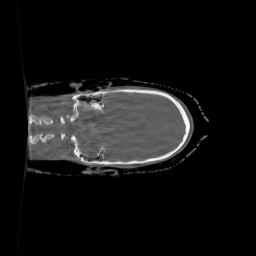}
\centerline{\scriptsize Few-view IR}
\centerline{\scriptsize 16.47~dB/0.9626}
\end{minipage}%
\hfill
\begin{minipage}[b]{0.19\linewidth}
\centering
\includegraphics[width=\linewidth]{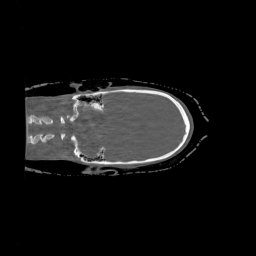}
\centerline{\scriptsize Few-view DIP}
\centerline{\scriptsize 21.10~dB/0.9885}
\end{minipage}%
\hfill
\begin{minipage}[b]{0.19\linewidth}
\centering
\includegraphics[width=\linewidth]{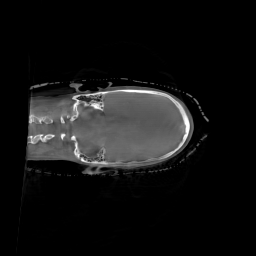}
\centerline{\scriptsize Limited-angle IR}
\centerline{\scriptsize 13.41~dB/0.9606}
\end{minipage}%
\hfill
\begin{minipage}[b]{0.19\linewidth}
\centering
\includegraphics[width=\linewidth]{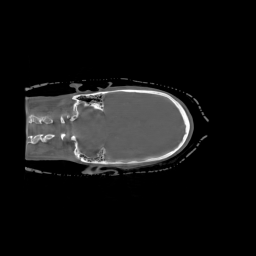}
\centerline{\scriptsize Limited-angle DIP}
\centerline{\scriptsize 19.16~dB/0.9914}
\end{minipage}%
\caption{Representative $256^3$ reconstruction results for the brain CT volume. The sparse-view setting uses 30 projections uniformly distributed over $360^\circ$; the limited-angle setting uses 120 projections acquired at $1^\circ$ intervals over $120^\circ$. Rows show axial, coronal, and sagittal slices; columns show the reference, sparse-view ASD-POCS, sparse-view DIP, limited-angle ASD-POCS, and limited-angle DIP. Each column is labeled below the third row, with reconstruction values reported as SNR in dB/SSIM.}
\label{fig:clinical256_brain}
\end{figure*}

Having evaluated network size and optimization speed, we assess the reconstruction quality of the selected compact DIP at $256^3$ using thoracic and brain CT volumes. Figures~\ref{fig:clinical256_chest} and~\ref{fig:clinical256_brain} provide representative illustrations of these experiments. The sparse-view setting uses 30 projections uniformly spaced over $360^\circ$. The limited-angle setting uses 120 projections at $1^\circ$ intervals within a $120^\circ$ acquisition range.

ASD-POCS starts from a zero volume and runs for 2500 outer iterations. DIP uses the four-level U-Net with 32 initial channels and 2500 RMSProp updates. For this comparison, the initial learning rate is $10^{-3}$ and is multiplied by 0.9 every 100 iterations.

Under these matched iteration budgets, DIP increases the final-iteration SNR relative to ASD-POCS by 2.33 and 7.07~dB for the thoracic sparse-view and limited-angle cases, respectively, and by 4.63 and 5.75~dB for the corresponding brain cases. The orthogonal slices show reduced streak artifacts and improved structural preservation, particularly under limited-angle acquisition. These results establish the feasibility and reconstruction quality of full-volume DIP at $256^3$, providing the structural estimate used to initialize the subsequent $512^3$ reconstruction.

\subsection{DIP-Guided High-Resolution Reconstruction}
\label{sec:resolution-extension}

We now evaluate the extension from $256^3$ DIP to $512^3$ reconstruction described in Section~\ref{sec:workflow}. Upsampling the DIP output and refining it with ASD-POCS avoids the eightfold increase in volume-sized feature maps required by direct network scaling. The two-stage strategy also combines the complementary strengths of the two methods: DIP suppresses streak artifacts caused by missing projections, while ASD-POCS provides stable physics-based refinement of measurement-supported details and mitigates network-specific artifacts. We first test whether DIP retains its reconstruction benefit on smaller grids, then assess the final $512^3$ output against direct ASD-POCS and direct full-resolution DIP.

The resolution experiment uses a modified Shepp--Logan phantom on $64^3$, $128^3$, $256^3$, and $512^3$ grids covering the same physical field of view. At each resolution, 30 projections are generated at $12^\circ$ intervals over $360^\circ$. ASD-POCS starts from a zero image and runs for 2500 outer iterations. DIP uses the same four-level U-Net with 32 initial channels and 2500 updates at every resolution. Reference images are used to compute SNR and SSIM, and all results are taken at the final iteration.

\begin{figure*}[htbp]
    \centering
    \begin{minipage}[b]{0.235\linewidth}
        \centering
        \includegraphics[width=\linewidth]{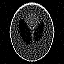}
        \centerline{(a) 4.99~dB / 0.5797}
    \end{minipage}%
    \hfill
    \begin{minipage}[b]{0.235\linewidth}
        \centering
        \includegraphics[width=\linewidth]{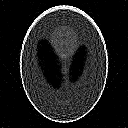}
        \centerline{(b) 8.44~dB / 0.6506}
    \end{minipage}%
    \hfill
    \begin{minipage}[b]{0.235\linewidth}
        \centering
        \includegraphics[width=\linewidth]{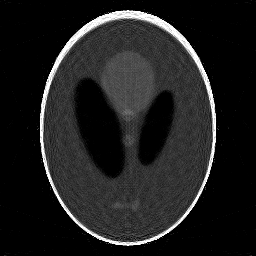}
        \centerline{(c) 14.19~dB / 0.7876}
    \end{minipage}%
    \hfill
    \begin{minipage}[b]{0.235\linewidth}
        \centering
        \includegraphics[width=\linewidth]{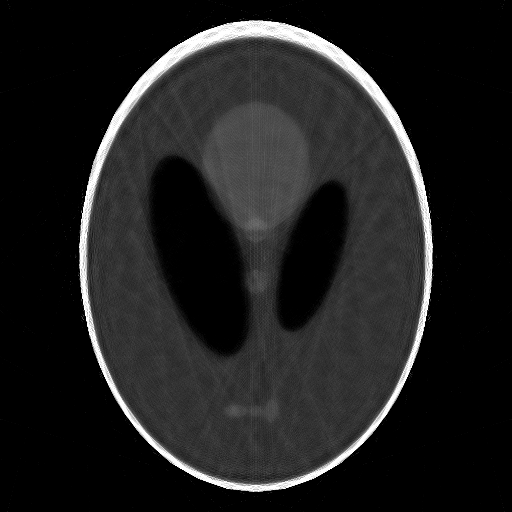}
        \centerline{(d) 15.54~dB / 0.8474}
    \end{minipage}

    \vspace{0.15cm}

    \begin{minipage}[b]{0.235\linewidth}
        \centering
        \includegraphics[width=\linewidth]{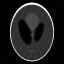}
        \centerline{(e) 21.04~dB / 0.8926}
    \end{minipage}%
    \hfill
    \begin{minipage}[b]{0.235\linewidth}
        \centering
        \includegraphics[width=\linewidth]{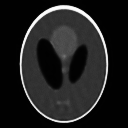}
        \centerline{(f) 21.85~dB / 0.8433}
    \end{minipage}%
    \hfill
    \begin{minipage}[b]{0.235\linewidth}
        \centering
        \includegraphics[width=\linewidth]{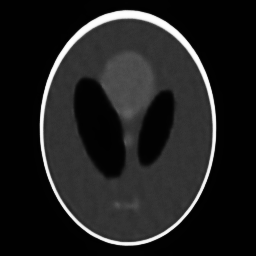}
        \centerline{(g) 21.45~dB / 0.8026}
    \end{minipage}%
    \hfill
    \begin{minipage}[b]{0.235\linewidth}
        \centering
        \includegraphics[width=\linewidth]{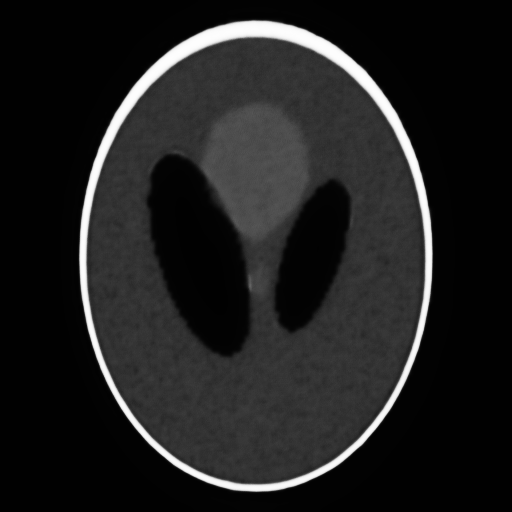}
        \centerline{(h) 20.87~dB / 0.7900}
    \end{minipage}
    \caption{ASD-POCS (a--d) and DIP (e--h) reconstructions of the three-dimensional modified Shepp--Logan phantom at $64^3$, $128^3$, $256^3$, and $512^3$. Both methods use 2500 iterations. Values below each image give SNR in dB and SSIM.}
    \label{fig:dip_resolution_images}
\end{figure*}

Across the four grids, ASD-POCS SNR increases from 4.99 to 15.54~dB, whereas DIP SNR remains between 20.87 and 21.85~dB (Fig.~\ref{fig:dip_resolution_images}). DIP suppresses the prominent streak artifacts seen in ASD-POCS, although its SSIM decreases from 0.8926 at $64^3$ to 0.7900 at $512^3$. The stable SNR indicates that the global reconstruction error changes little with grid size, whereas the decreasing SSIM suggests that local neural-network-specific artifacts become more pronounced as the DIP resolution increases. This result supports using DIP at reduced resolution to obtain an artifact-suppressed structural estimate and relying on subsequent ASD-POCS refinement to mitigate local network-specific artifacts at the target resolution.

\begin{figure}[htbp]
   \begin{minipage}[b]{0.3\linewidth}
        \centering
        \includegraphics[width=\linewidth]{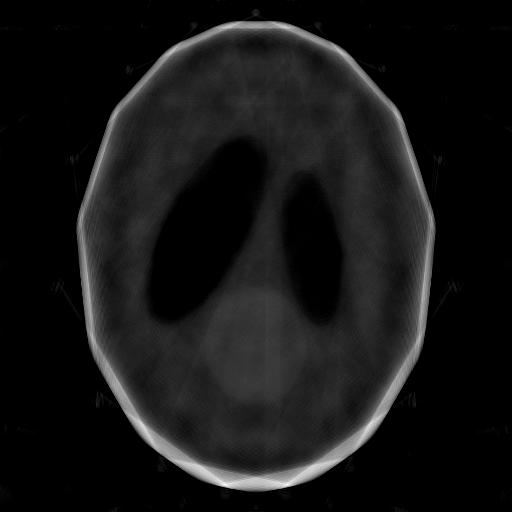}
        \centerline{10.65~dB}
    \end{minipage}
    \begin{minipage}[b]{0.3\linewidth}
        \centering
        \includegraphics[width=\linewidth]{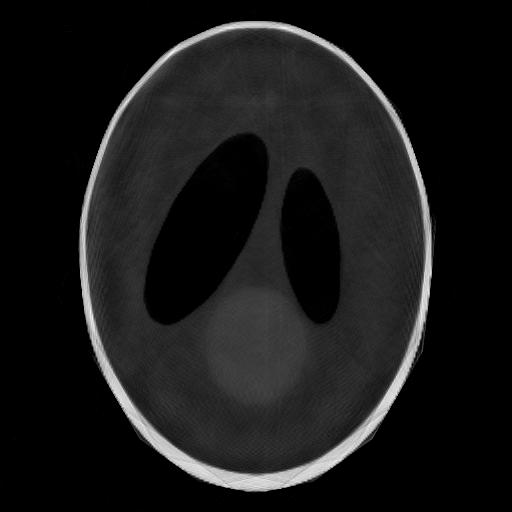}
        \centerline{11.31~dB}
    \end{minipage}
    \begin{minipage}[b]{0.3\linewidth}
        \centering
        \includegraphics[width=\linewidth]{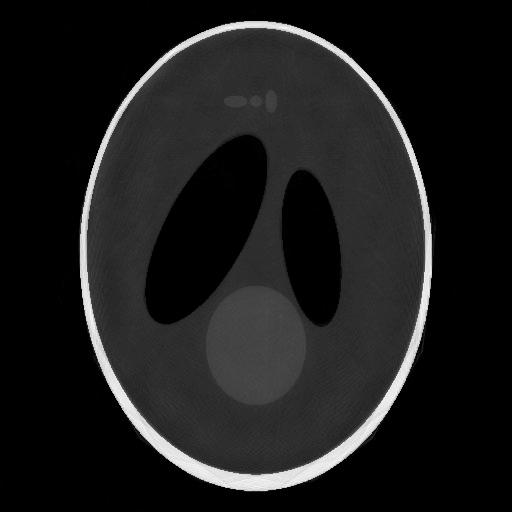}
        \centerline{14.98~dB}
    \end{minipage}

    \vspace{0.2cm}
    \begin{minipage}[b]{0.3\linewidth}
        \centering
        \includegraphics[width=\linewidth]{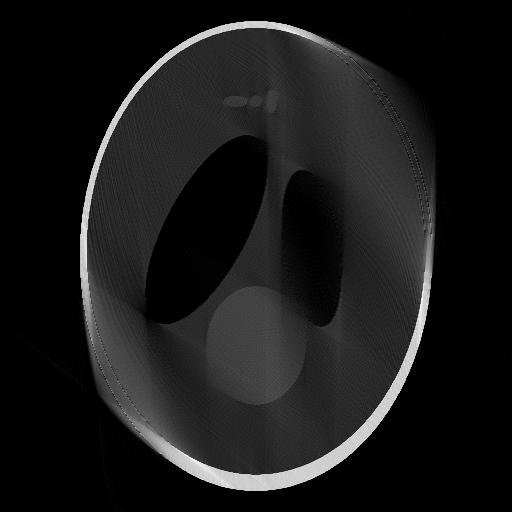}
        \centerline{10.54~dB}
        \centerline{ASD-POCS}
    \end{minipage}
    \begin{minipage}[b]{0.3\linewidth}
        \centering
        \includegraphics[width=\linewidth]{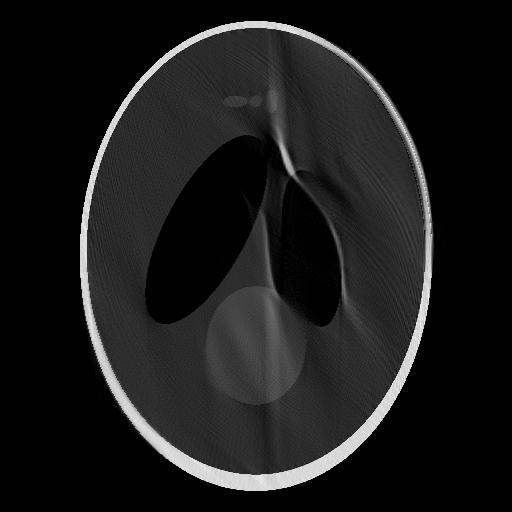}
        \centerline{15.36~dB}
        \centerline{DIP}
    \end{minipage}
    \begin{minipage}[b]{0.3\linewidth}
        \centering
        \includegraphics[width=\linewidth]{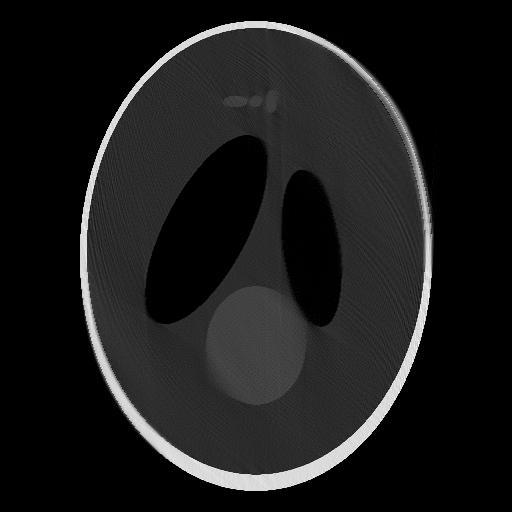}
        \centerline{17.98~dB}
        \centerline{3D-DIP}
    \end{minipage}
    \caption{Direct ASD-POCS, direct full-resolution DIP, and the proposed 3D-DIP (left to right) under sparse-view (top) and limited-angle (bottom) acquisition. Sparse-view data comprise 30 projections uniformly distributed over $360^\circ$; limited-angle data comprise 120 projections acquired at $1^\circ$ intervals over $120^\circ$. Values below the images are SNR in dB.}
    \label{fig:method_comparison}
\end{figure}

Figure~\ref{fig:method_comparison} compares direct ASD-POCS, direct full-resolution DIP, and the proposed 3D-DIP. The latter uses 2500 DIP updates at $256^3$, trilinear interpolation, and 1000 ASD-POCS iterations at $512^3$. Direct ASD-POCS uses the same full-resolution iteration budget, while direct DIP performs 2500 updates directly on the $512^3$ reconstruction grid. Sparse-view acquisition uses 30 projections over $360^\circ$, and limited-angle acquisition uses 120 projections at $1^\circ$ intervals over a $120^\circ$ range.

Direct ASD-POCS exhibits prominent streak artifacts under both acquisition conditions. Direct full-resolution DIP reduces these artifacts but retains local distortions. The proposed 3D-DIP gives the highest SNR: 14.98~dB for sparse-view reconstruction and 17.98~dB for limited-angle reconstruction, compared with 10.65 and 10.54~dB for ASD-POCS and 11.31 and 15.36~dB for direct DIP, respectively. The images show improved structural recovery with the proposed framework.

Direct DIP reconstruction on a $512\times512\times512$ grid exceeds the memory capacity of current single-GPU hardware because the complete network state and intermediate tensors must be retained at the full target resolution. To obtain the direct full-resolution DIP comparison, we therefore used a segmented implementation. Network states, activations, and gradients were exchanged through host memory and disk, and projector operations were evaluated sequentially. This implementation was used to obtain the controlled comparison in Fig.~\ref{fig:method_comparison}; its data-transfer requirements precluded its use in the broader clinical-volume evaluation.

These comparisons support extending the $256^3$ structural estimate to a $512^3$ reconstruction through ASD-POCS refinement. The interpretation in terms of complementary priors is developed in Section~\ref{sec:discussion}.

\subsection{Evaluation under Varying Angular Sampling Conditions}
\label{sec:clinical-volumes}
We next examine how the complete $512^3$ framework performs as angular information becomes increasingly incomplete. The evaluation includes thoracic and brain CT volumes, extending the controlled comparison in Section~\ref{sec:resolution-extension} to anatomical structures and a broader range of sampling conditions. Sparse-view data contain 15, 30, 60, 120, or 240 views uniformly distributed over $360^\circ$. The angular-coverage study uses $1^\circ$ sampling intervals over ranges of $90^\circ$, $120^\circ$, $150^\circ$, $180^\circ$, and $360^\circ$, with the last serving as the full-angular-coverage reference.

We compare 3D-DIP with direct ASD-POCS using the same target grid, projection data, and 1000-iteration ASD-POCS budget. Direct full-resolution DIP is restricted to the controlled comparison in Section~\ref{sec:resolution-extension} because of its memory and data-transfer requirements.

\begin{figure}[htbp]
    \centering
    \setstretch{1}
    \centering
    \begin{minipage}[b]{0.85\linewidth}
        \centering
        \includegraphics[width=\linewidth]{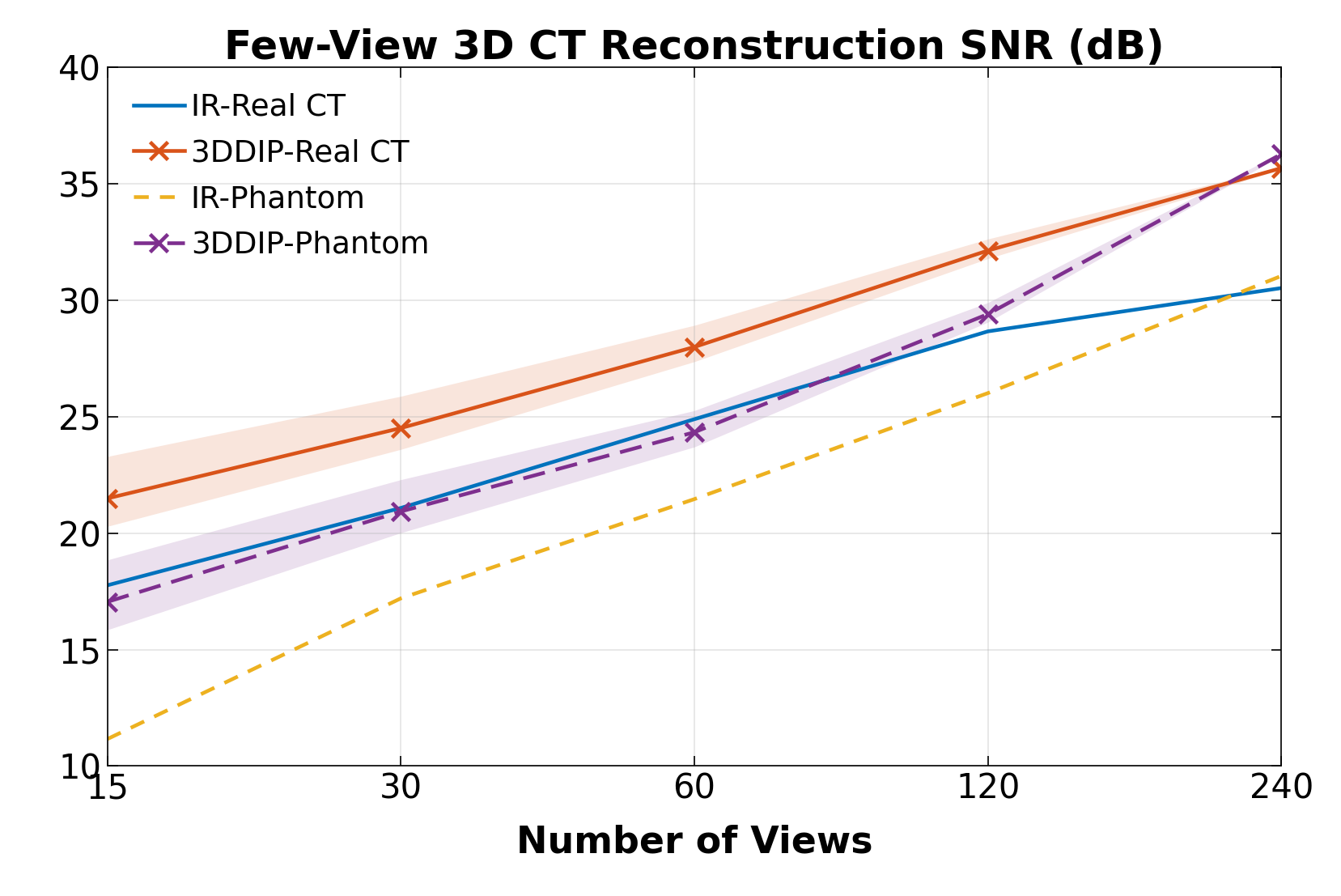}
        \centerline{(a) Sparse-view CT}
    \end{minipage}

    \vspace{0.2cm}
    \begin{minipage}[b]{0.85\linewidth}
        \centering
        \includegraphics[width=\linewidth]{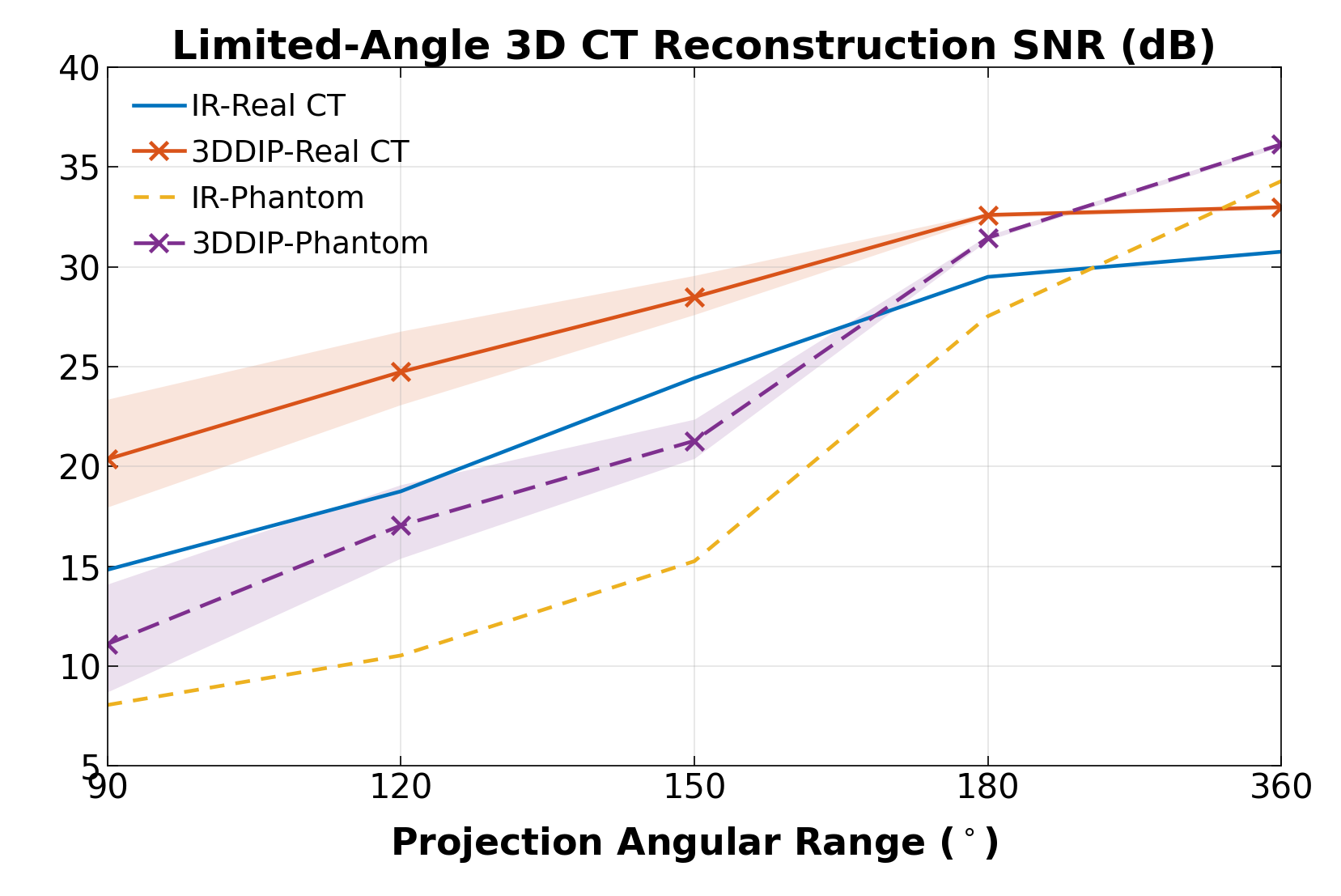}
        \centerline{(b) Limited-angle CT}
    \end{minipage}
    \caption{Reconstruction SNR for synthetic phantoms and clinical CT volumes under (a) sparse-view acquisition with 15--240 views over $360^\circ$ and (b) angular coverage from $90^\circ$ to $360^\circ$ with $1^\circ$ sampling intervals. The central 3D-DIP curves show the mean SNR, and the shaded regions show the observed ranges across repeated reconstructions. In the legends, IR denotes ASD-POCS, 3DDIP denotes the proposed 3D-DIP, and Real CT denotes reconstructions of clinical CT volumes.}
    \label{fig:real_snr}
\end{figure}

Figure~\ref{fig:real_snr} summarizes SNR as the number of views or angular coverage changes. When only a small number of views or a narrow angular range is available, the 3D-DIP results exhibit greater variability because the severely incomplete measurements provide weaker constraints on the reconstruction. As the number of views or angular coverage increases, the error bands progressively narrow, indicating that 3D-DIP becomes more stable as additional projection information is provided. Although ASD-POCS also benefits from increased angular information, the lower bounds of the 3D-DIP results remain above the corresponding ASD-POCS results for both the clinical CT volumes and the phantoms throughout the tested range. These results show that the low-resolution DIP estimate provides useful structural guidance across different degrees of angular incompleteness, with increasingly stable performance as the available projection information grows.

\begin{figure}[htbp]
    \centering
    \setstretch{1}
    \begin{minipage}[b]{0.205\linewidth}
        \centering
        \includegraphics[width=\linewidth]{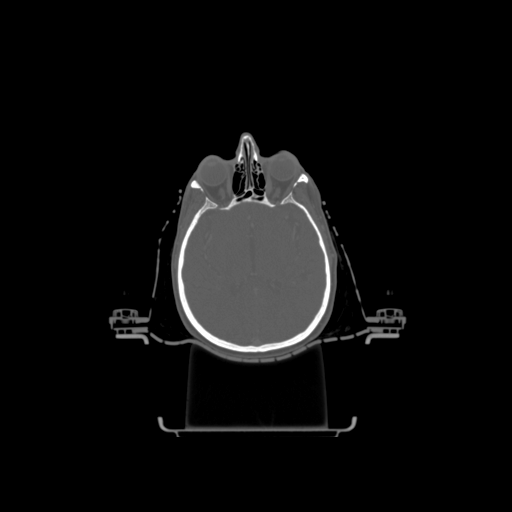}
    \end{minipage}%
    \begin{minipage}[b]{0.205\linewidth}
        \centering
        \includegraphics[width=\linewidth]{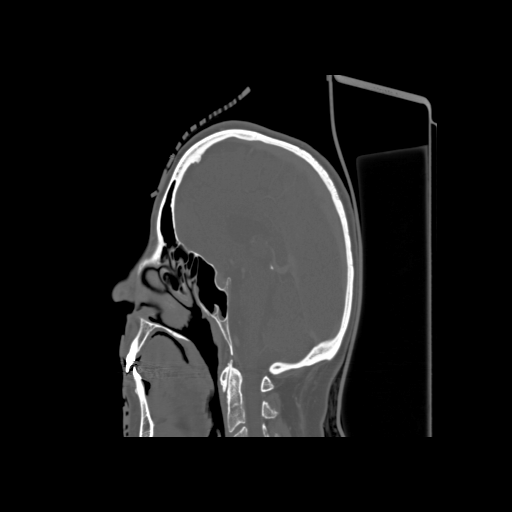}
    \end{minipage}%
    \begin{minipage}[b]{0.205\linewidth}
        \centering
        \includegraphics[width=\linewidth]{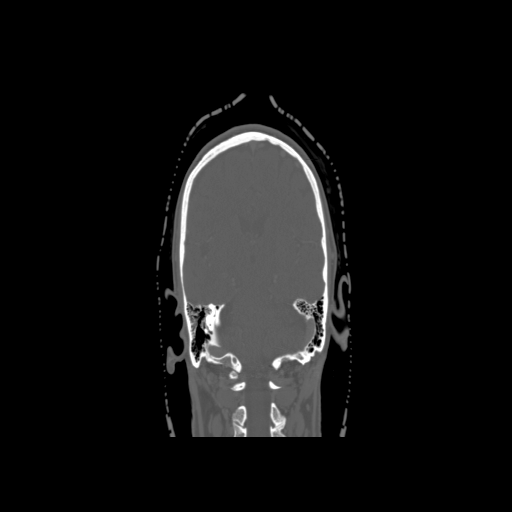}
    \end{minipage}
    \centerline{Reference volume (reconstruction labels: SNR/SSIM)}

    \begin{minipage}[b]{0.205\linewidth}
        \centering
        \includegraphics[width=\linewidth]{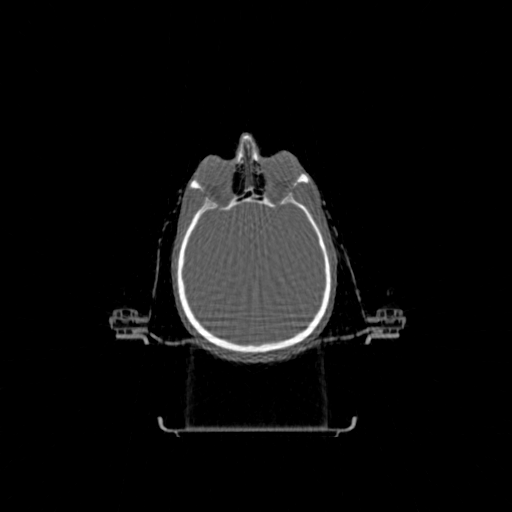}
    \end{minipage}%
    \begin{minipage}[b]{0.205\linewidth}
        \centering
        \includegraphics[width=\linewidth]{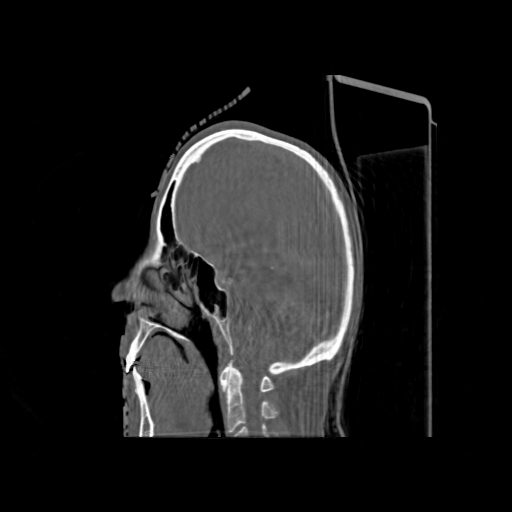}
    \end{minipage}%
    \begin{minipage}[b]{0.205\linewidth}
        \centering
        \includegraphics[width=\linewidth]{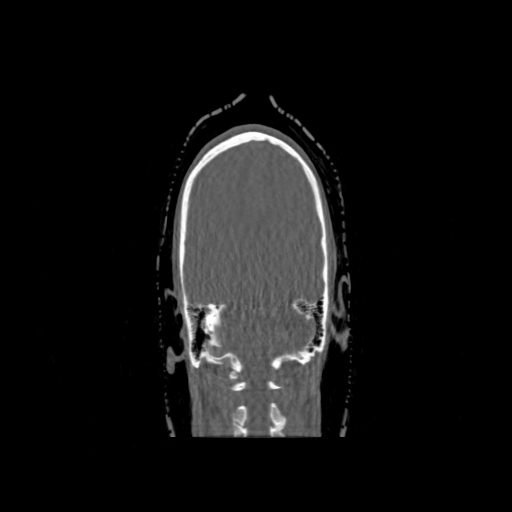}
    \end{minipage}
    \centerline{Sparse-view ASD-POCS  19.84~dB/0.9818}   
    
    \begin{minipage}[b]{0.205\linewidth}
        \centering
        \includegraphics[width=\linewidth]{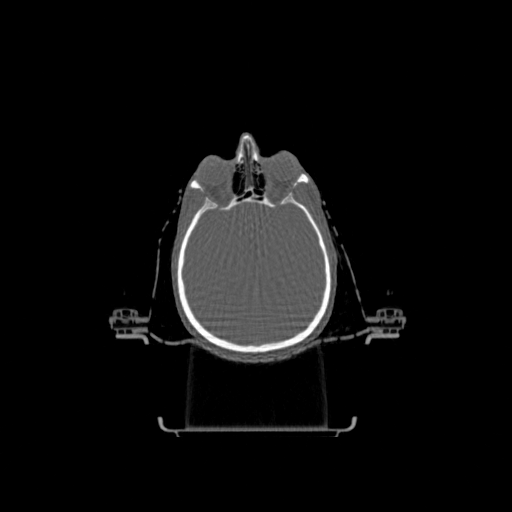}
    \end{minipage}%
    \begin{minipage}[b]{0.205\linewidth}
        \centering
        \includegraphics[width=\linewidth]{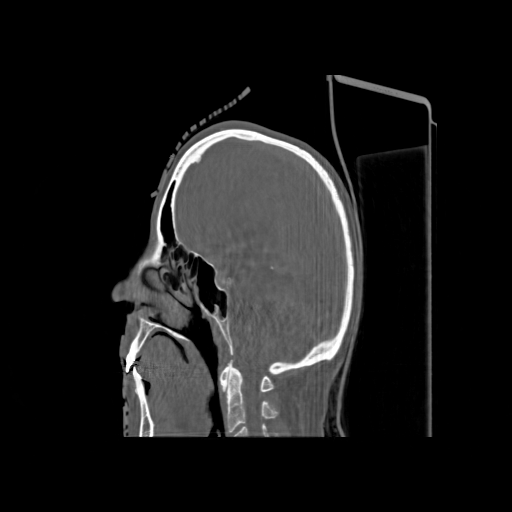}
    \end{minipage}%
    \begin{minipage}[b]{0.205\linewidth}
        \centering
        \includegraphics[width=\linewidth]{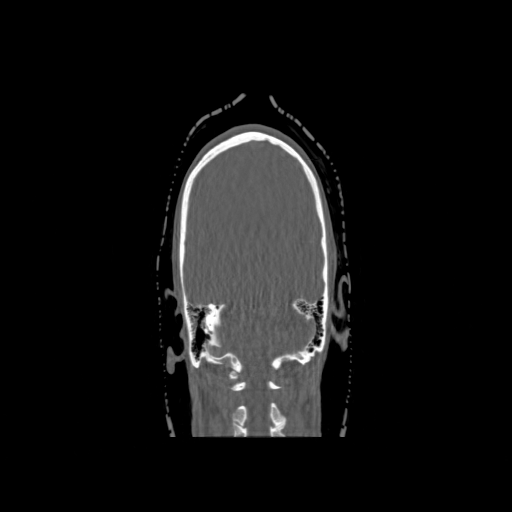}
    \end{minipage}
    \centerline{Sparse-view 3D-DIP 23.67~dB/0.9904}

    \begin{minipage}[b]{0.205\linewidth}
        \centering
        \includegraphics[width=\linewidth]{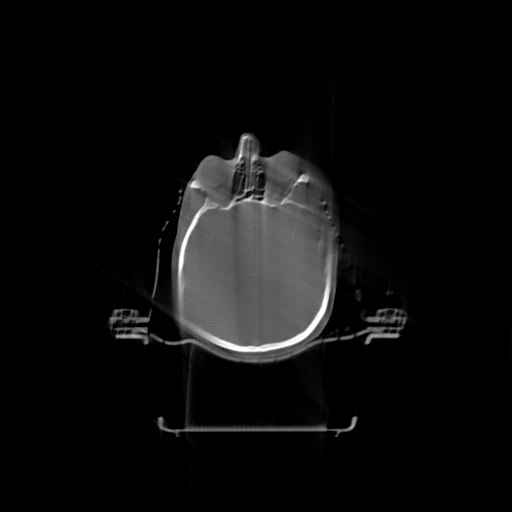}
    \end{minipage}%
    \begin{minipage}[b]{0.205\linewidth}
        \centering
        \includegraphics[width=\linewidth]{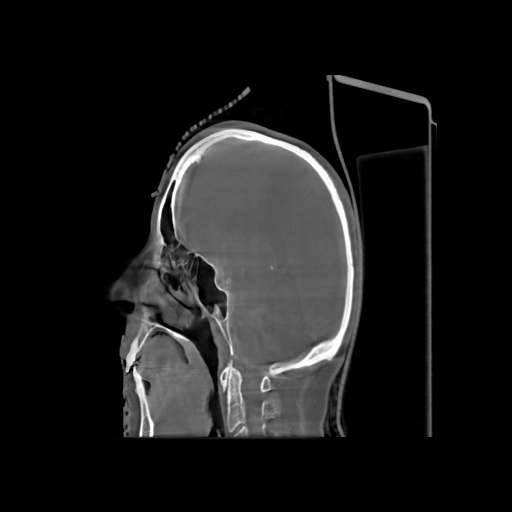}
    \end{minipage}%
    \begin{minipage}[b]{0.205\linewidth}
        \centering
        \includegraphics[width=\linewidth]{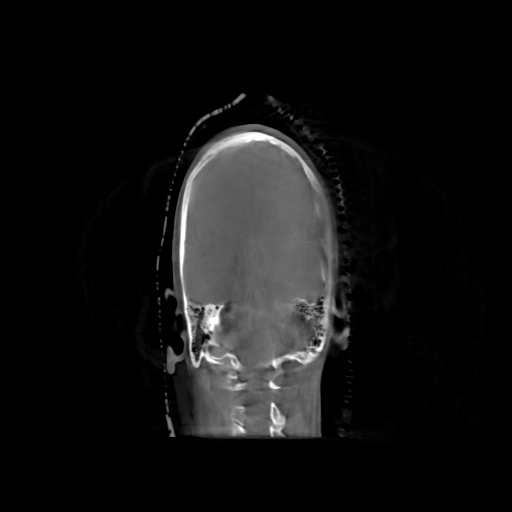}
    \end{minipage}
    \centerline{Limited-angle ASD-POCS 12.10~dB/0.9615}

    \begin{minipage}[b]{0.205\linewidth}
        \centering
        \includegraphics[width=\linewidth]{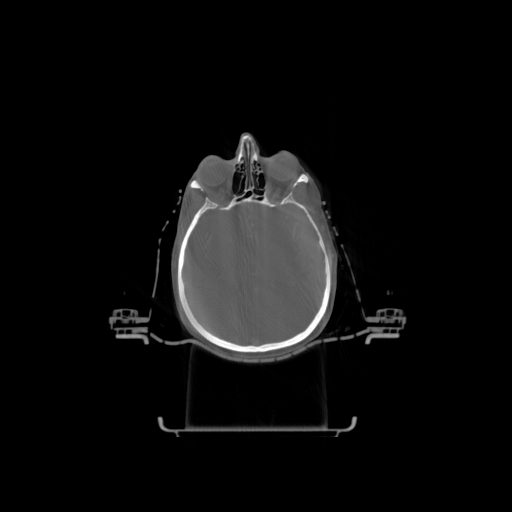}
    \end{minipage}%
    \begin{minipage}[b]{0.205\linewidth}
        \centering
        \includegraphics[width=\linewidth]{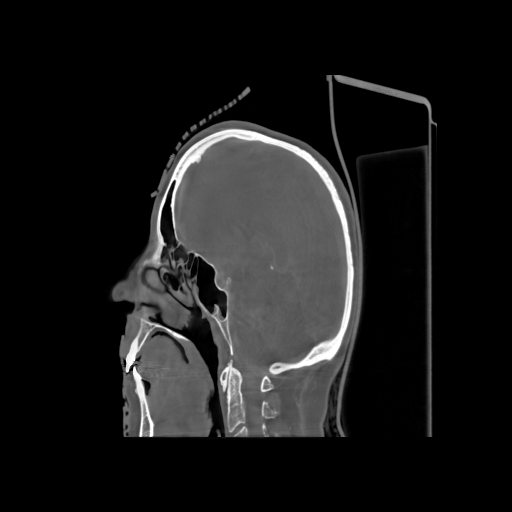}
    \end{minipage}%
    \begin{minipage}[b]{0.205\linewidth}
        \centering
        \includegraphics[width=\linewidth]{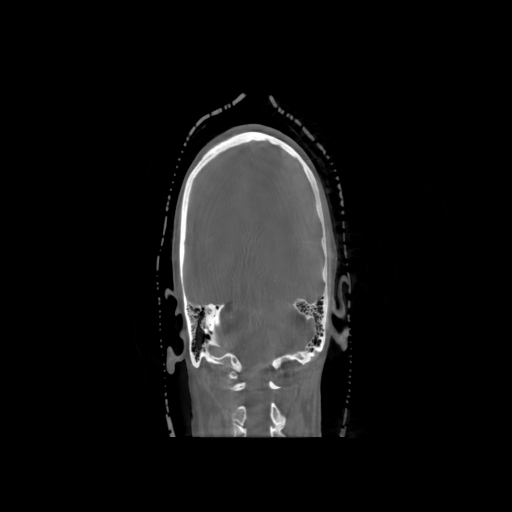}
    \end{minipage}
    \centerline{Limited-angle 3D-DIP 19.01~dB/0.9908}
    
    \caption{Representative reconstruction results from simulated projections of a brain CT volume using ASD-POCS and the proposed 3D-DIP. The sparse-view setting uses 30 projections uniformly distributed over $360^{\circ}$; the limited-angle setting uses 120 projections acquired at $1^\circ$ intervals over $120^{\circ}$. Rows show the reference volume, sparse-view ASD-POCS, sparse-view 3D-DIP, limited-angle ASD-POCS, and limited-angle 3D-DIP; columns show three orthogonal anatomical views. Reconstruction values are reported as SNR in dB/SSIM.}
    \label{fig:brainCT}
\end{figure}

\begin{figure}[htbp]
    \centering
    \setstretch{1}
    \begin{minipage}[b]{0.205\linewidth}
        \centering
        \includegraphics[width=\linewidth]{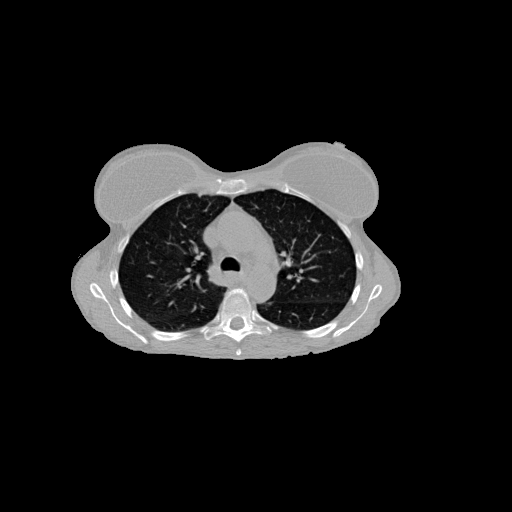}
    \end{minipage}%
    \begin{minipage}[b]{0.205\linewidth}
        \centering
        \includegraphics[width=\linewidth]{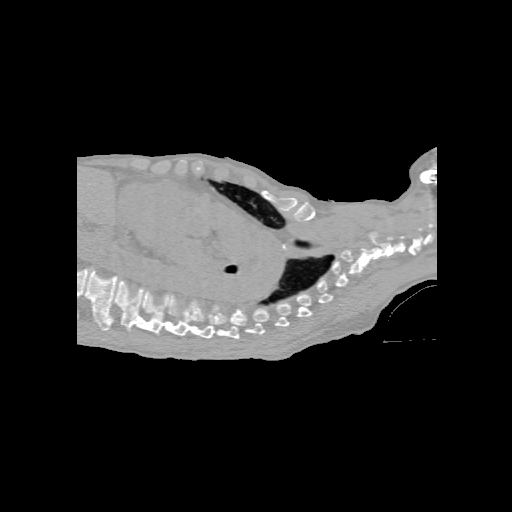}
    \end{minipage}%
    \begin{minipage}[b]{0.205\linewidth}
        \centering
        \includegraphics[width=\linewidth]{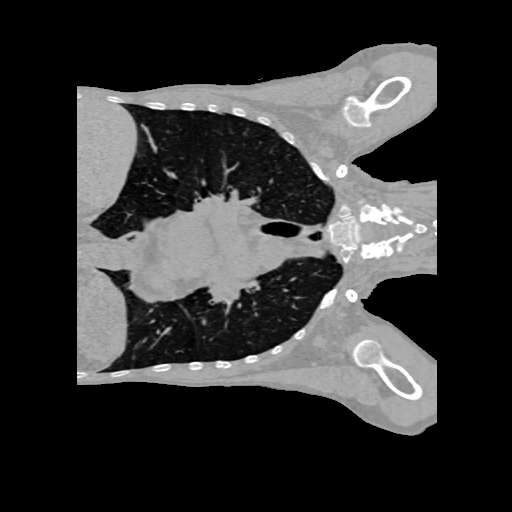}
    \end{minipage}
    \centerline{Reference volume (reconstruction labels: SNR/SSIM)}

    \begin{minipage}[b]{0.205\linewidth}
        \centering
        \includegraphics[width=\linewidth]{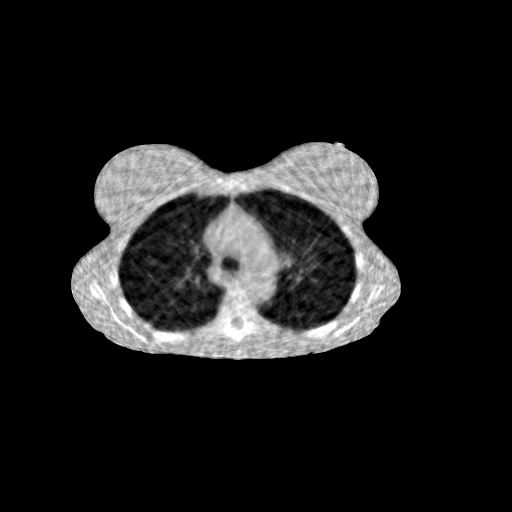}
    \end{minipage}%
    \begin{minipage}[b]{0.205\linewidth}
        \centering
        \includegraphics[width=\linewidth]{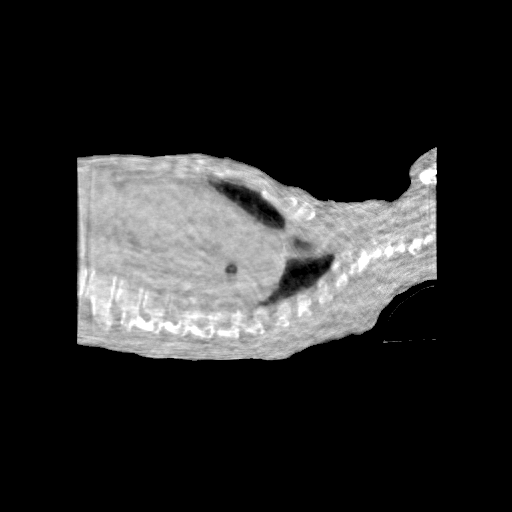}
    \end{minipage}%
    \begin{minipage}[b]{0.205\linewidth}
        \centering
        \includegraphics[width=\linewidth]{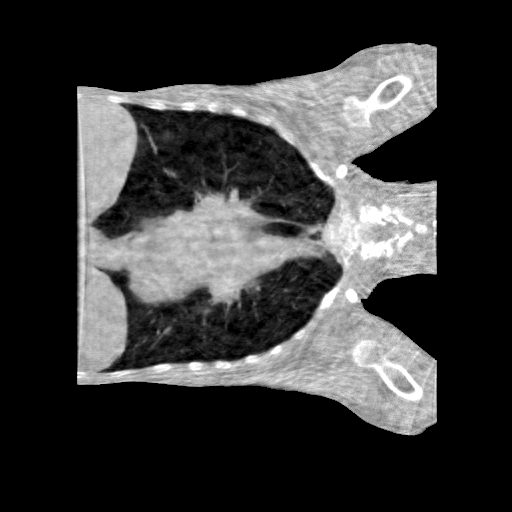}
    \end{minipage}
    \centerline{Sparse-view ASD-POCS 21.04~dB/0.9812}

    \begin{minipage}[b]{0.205\linewidth}
        \centering
        \includegraphics[width=\linewidth]{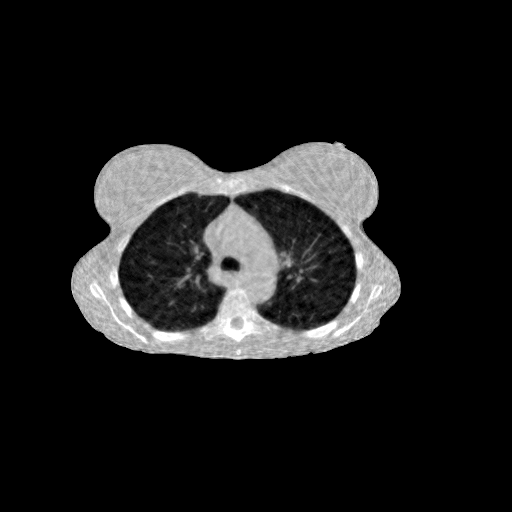}
    \end{minipage}%
    \begin{minipage}[b]{0.205\linewidth}
        \centering
        \includegraphics[width=\linewidth]{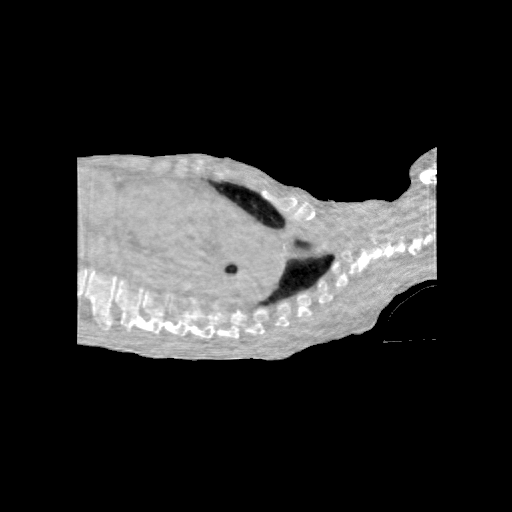}
    \end{minipage}%
    \begin{minipage}[b]{0.205\linewidth}
        \centering
        \includegraphics[width=\linewidth]{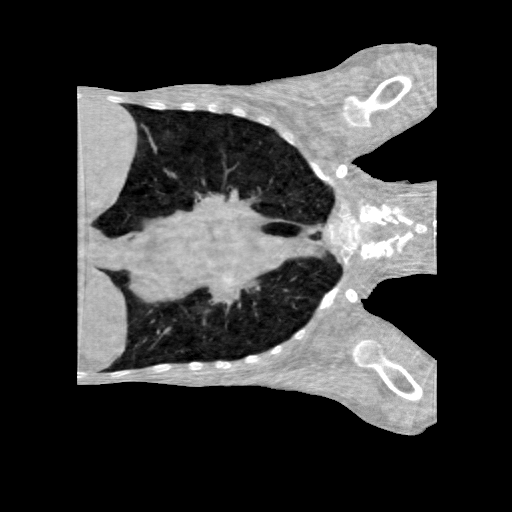}
    \end{minipage}
    \centerline{Sparse-view 3D-DIP 24.17~dB/0.9891}

    \begin{minipage}[b]{0.205\linewidth}
        \centering
        \includegraphics[width=\linewidth]{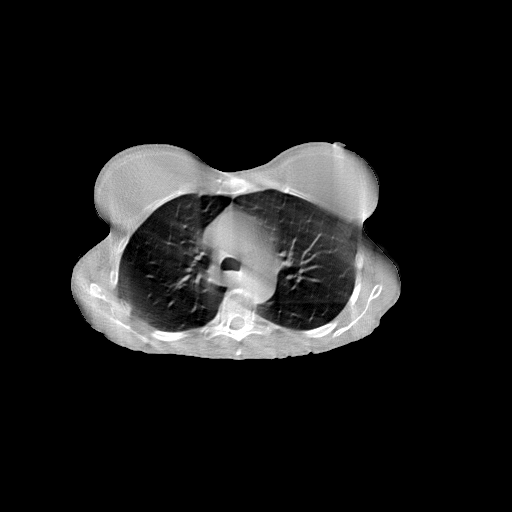}
    \end{minipage}%
    \begin{minipage}[b]{0.205\linewidth}
        \centering
        \includegraphics[width=\linewidth]{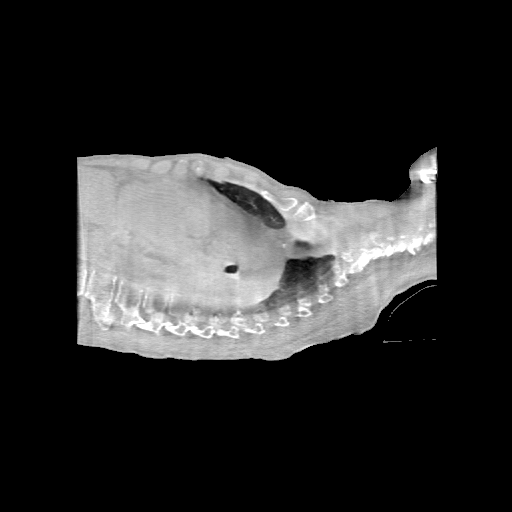}
    \end{minipage}%
    \begin{minipage}[b]{0.205\linewidth}
        \centering
        \includegraphics[width=\linewidth]{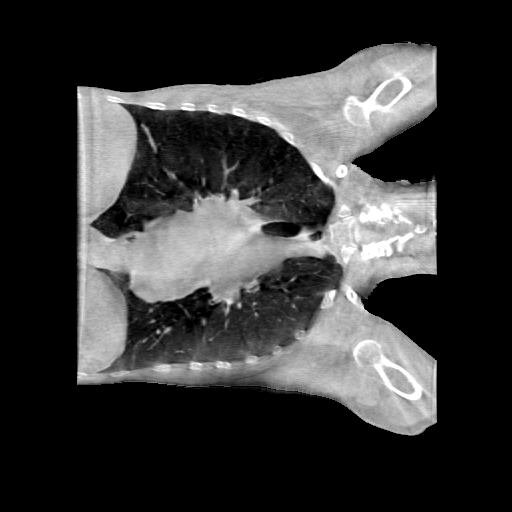}
    \end{minipage}
    \centerline{Limited-angle ASD-POCS 17.69~dB/0.9724}

    \begin{minipage}[b]{0.205\linewidth}
        \centering
        \includegraphics[width=\linewidth]{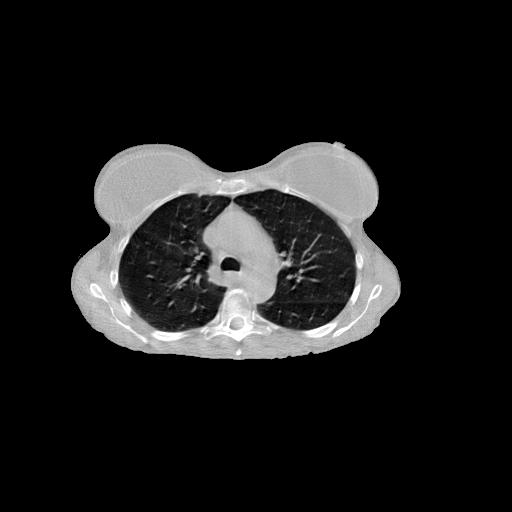}
    \end{minipage}%
    \begin{minipage}[b]{0.205\linewidth}
        \centering
        \includegraphics[width=\linewidth]{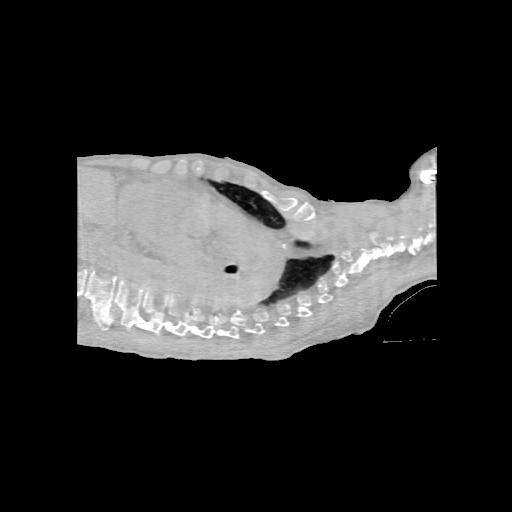}
    \end{minipage}%
    \begin{minipage}[b]{0.205\linewidth}
        \centering
        \includegraphics[width=\linewidth]{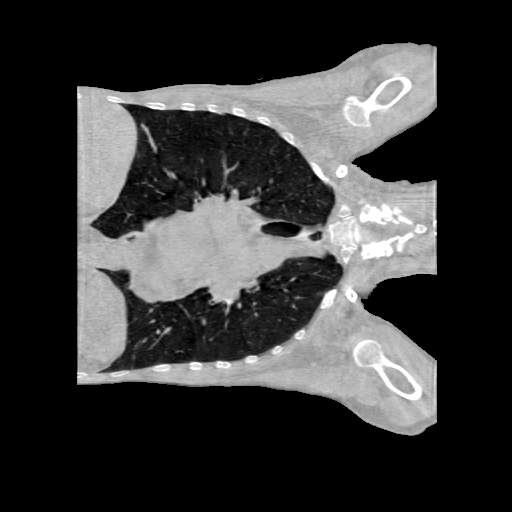}
    \end{minipage}
    \centerline{Limited-angle 3D-DIP 25.26~dB/0.9925}
    
    \caption{Representative reconstruction results from simulated projections of a thoracic CT volume using ASD-POCS and the proposed 3D-DIP. The sparse-view setting uses 30 projections uniformly distributed over $360^{\circ}$; the limited-angle setting uses 120 projections acquired at $1^\circ$ intervals over $120^{\circ}$. Rows show the reference volume, sparse-view ASD-POCS, sparse-view 3D-DIP, limited-angle ASD-POCS, and limited-angle 3D-DIP; columns show three orthogonal anatomical views. Reconstruction values are reported as SNR in dB/SSIM.}
    \label{fig:chestCT}
\end{figure}

Figures~\ref{fig:brainCT} and~\ref{fig:chestCT} show orthogonal slices from the brain and thoracic reconstructions. The proposed 3D-DIP reduces streak and missing-angle artifacts while preserving anatomical structure. In the brain example, SNR increases from 19.84 to 23.67~dB for sparse-view CT and from 12.10 to 19.01~dB for limited-angle CT, corresponding to gains of 3.83 and 6.91~dB. The thoracic gains are 3.13 and 7.57~dB, respectively.

\subsection{GPU Memory and Runtime Cost}
\label{sec:computational-cost}
The proposed 3D-DIP requires approximately 18~GB of peak GPU memory on an RTX~4090. The measured runtime for 2500 DIP updates at $256^3$ is approximately 1.3 minutes with 30 projections and 1.7 minutes with 120 projections. These measurements are consistent with the mean pipelined per-iteration times reported in Table~\ref{tab:pipeline_time}. The subsequent $512^3$ ASD-POCS stage requires approximately 12 minutes for 1000 outer iterations, while the cost of trilinear interpolation is negligible. The combined optimization time is therefore approximately 13.3--13.7 minutes. Relative to direct ASD-POCS with the same full-resolution iteration budget, the low-resolution DIP stage introduces an additional cost of approximately 1.3--1.7 minutes.

Direct GPU-resident DIP optimization at $512^3$ requires substantially more memory because volumetric activations and gradients must be retained at the full target resolution. This requirement exceeds the capacity of current single-GPU hardware, whereas the proposed low-resolution DIP-guided workflow remains within 18~GB of GPU memory.

\section{Discussion}
\label{sec:discussion}
The findings suggest that the resolution at which a neural prior is formed need not be the resolution of the final reconstruction. In 3D-DIP, the network resolves the principal missing-data ambiguity on a computationally feasible $256^3$ grid, whereas the subsequent $512^3$ stage restores measurement-supported detail without allocating full-resolution network features. This separation is what allows the method to progress from practical full-volume DIP to high-resolution DIP-guided reconstruction. Network simplification and pipelined optimization make the prior-generation stage efficient, while ASD-POCS provides a comparatively inexpensive route to the target grid.

This role of the reduced-resolution reconstruction distinguishes 3D-DIP from multiresolution IR that retains the same model and regularizer across grid levels~\cite{cao2016multiresolution}. A coarse grid in conventional multiresolution IR primarily improves numerical efficiency; it does not introduce a qualitatively different constraint on the solution. Here, resolution transfer also marks a transition between regularization mechanisms. The reduced-resolution network contributes a nonlinear architectural prior, whereas full-resolution ASD-POCS imposes physics-based data consistency, nonnegativity, and explicit TV regularization. The method therefore uses the lower-resolution reconstruction to select a plausible solution family before refining that solution on the target grid.

The two stages also have complementary artifact behavior. DIP favors spatially coherent images and is effective at suppressing streaks caused by missing projections, but prolonged network optimization can introduce local network-specific patterns. ASD-POCS alone cannot resolve the same missing-data ambiguity as effectively, yet its forward-model updates and explicit image-domain regularization can stabilize the DIP-derived candidate and attenuate local variations. Because the inverse problem is highly ill posed, the initialization influences which data-consistent candidate is reached. The upsampled DIP reconstruction accordingly serves as more than a coarse numerical starting point: it transfers a nonlinear structural choice into the full-resolution optimization, after which ASD-POCS refines rather than recreates the global structure.

The pipeline has an implication beyond operation scheduling. It combines a one-step-delayed image-domain gradient with the current network Jacobian and therefore introduces a small perturbation into the DIP optimization trajectory. DIP is known to recover dominant image structure before fitting noise or model mismatch~\cite{ulyanov2018deep,cascarano2022first}; the delayed update may mildly oppose this late-stage fitting. This interpretation is consistent with the absence of an observed accuracy loss and the small final-SNR improvement in the pipeline experiments. The principal established benefit remains computational: projector work can be hidden behind network computation when their execution times are suitably balanced.

ASD-POCS was selected as the primary comparator because 3D-DIP remains a scan-specific iterative reconstruction method: it replaces a directly parameterized image with a nonlinear network parameterization but optimizes that representation from the measured projections without pretrained weights. Comparison with conventional IR therefore directly tests whether the nonlinear architectural prior improves solution selection under the same acquisition data. A matched comparison with pretrained reconstruction models presents a different requirement. Such models depend strongly on the size, quality, and representativeness of their training data, yet publicly available high-quality volumetric CT datasets with matched three-dimensional projection data remain limited. Differences in training distribution, acquisition geometry, and preprocessing could consequently confound a direct comparison in the present study.

The thoracic and brain examples extend the evaluation from synthetic phantoms to anatomical structures under multiple acquisition conditions. Future work will collect additional high-quality volumetric CT data and acquired raw projections to evaluate performance across scanners, protocols, anatomies, and patient populations. A larger dataset will also enable comparisons with pretrained reconstruction models under matched projection geometry, output resolution, noise conditions, and hardware. These studies can clarify the relative benefits of scan-specific DIP optimization and population-trained priors while extending the present validation beyond the evaluated cases.

\section{Conclusion}
\label{sec:conclusion}
We presented 3D-DIP, a training-free framework for volumetric CT reconstruction under sparse-view and limited-angle acquisition. A compact 3D network and delayed-gradient pipeline enable practical full-volume DIP reconstruction at $256^3$. The resulting structural estimate is then transferred to a $512^3$ grid and refined with ASD-POCS, combining a nonlinear architectural prior with physics-based data consistency, nonnegativity, and explicit TV regularization.

Across the phantom, brain, and thoracic experiments, 3D-DIP consistently reduced missing-projection artifacts and improved reconstruction SNR relative to direct ASD-POCS. The complete framework operated within approximately 18~GB of GPU memory and introduced only a modest runtime increase over full-resolution ASD-POCS. These results demonstrate that a neural prior need not be optimized directly at the final reconstruction resolution. Separating reduced-resolution prior generation from full-resolution physics-based refinement provides a practical route to high-resolution, training-free three-dimensional CT reconstruction and may be extended to other DIP architectures and computationally demanding inverse problems.


\section*{Disclosures}
The authors declare there are no financial interests, commercial affiliations,
or other potential conflicts of interest that have influenced the objectivity
of this research or the writing of this paper.

\section*{Code and Data Availability}
The source code associated with this study is publicly available at
\url{https://github.com/Flocculus/3D-DIP}.
The clinical CT data used in this study are not publicly available due to
patient privacy and institutional restrictions and cannot be shared outside
Stony Brook University Hospital.

\section*{Acknowledgments}
This work was supported in part by:
\begin{itemize}
\item NIH STTR Phase II grant R42EB026401
\item NIH/NCI, SBIR Phase II Contract 75N91022C00055
\item Stony Brook's AI Innovation Institute Awards
\end{itemize}


\bibliography{refs}   
\bibliographystyle{spiejour}   



\listoffigures
\listoftables

\end{spacing}
\end{document}